**Comparing Methods of Characterizing Energetic Disorder in Organic Solar Cells**

*Paula Hartnagel, Sandheep Ravishankar, Benjamin Klingebiel, Oliver Thimm and Thomas Kirchartz\**

P. Hartnagel, S. Ravishankar, B. Klingebiel, O. Thimm
IEK5-Photovoltaik, Forschungszentrum Jülich, 52425 Jülich, Germany

T. Kirchartz
IEK5-Photovoltaik, Forschungszentrum Jülich, 52425 Jülich, Germany
Faculty of Engineering and CENIDE, University of Duisburg-Essen, Carl-Benz-Str. 199, 47057 Duisburg, Germany
E-mail: t.kirchartz@fz-juelich.de

Abstract: Energetic disorder has been known for decades to limit the performance of structurally disordered semiconductors such as amorphous silicon and organic semiconductors. However, in the past years, high performance organic solar cells have emerged showing a continuously reduced amount of energetic disorder. While searching for future high efficiency material systems, it is therefore important to correctly characterize this energetic disorder. While there are several techniques in literature, the most common approaches to probe the density of defect states are using optical excitation as in external quantum efficiency measurements or sequential filling of the tail states by applying an external voltage as in admittance spectroscopy. A metanalysis of available literature as well as our experiments using four characterization techniques on two material systems reveal that electrical, voltage-dependent measurements frequently yield higher values of energetic disorder than optical measurements. With drift-diffusion simulations, we demonstrate that the approaches probe different energy ranges of the subband-gap density of states. We further explore the limitations of the techniques and find that extraction of information from a capacitance-voltage curve can be inhibited by an internal series resistance. Thereby, we explain the discrepancies between measurements techniques with sensitivity to different energy ranges and electronic parameters.

## 1. Introduction

Amorphous inorganic semiconductors have been studied over decades for photovoltaic applications but were eventually discarded, because of poor performance that hardly exceeded the 10% mark.[1-3] The substantially reduced performance of e.g. amorphous silicon relative to crystalline silicon is to a major degree caused by increased charge trapping and recombination due to energetic disorder evoked by the structural disorder of the amorphous materials.[4-6] In the world of organic semiconductors, disorder has also been frequently considered to be a major obstacle towards reaching higher efficiencies.[7-9] For a long time during the development of organic photovoltaics, efficiencies were struggling to reach 10% leading to a seemingly similar situation as with amorphous silicon solar cells.[10] Over the last several years, however, new polymers and especially new acceptor molecules have completely changed the situation. By now, organic solar cell efficiencies exceed 19% in single-junction devices[11-15] and 20% for tandem organic solar cells,[16] thereby reaching efficiencies nearly twice that of typical amorphous Si solar cells. Thus, while structural and energetic disorder will certainly still be present in current state-of-the-art organic solar cells, either the degree or the impact of the disorder has to be substantially reduced relative to amorphous silicon. Hence, a closer look at energetic disorder and its impact on device performance for current generations of organic solar cells is needed. To achieve this goal, it is crucial to use sensible parameters to quantify disorder, to use reliable measurement methods and to be aware of potential limitations of the methods.

Energetic disorder as observed e.g. in absorption and emission measurements is explained by two significantly different physical concepts that are usually referred to as static and dynamic disorder. Static disorder originates from a broadened density of states that is caused by structural disorder of e.g. polymer chains.[17] Dynamic disorder is caused by a combination of two effects: The first is the presence of a non-zero reorganization energy, i.e. a displacement of the vibrational ground state of the electronically excited state relative to the electronic ground state. The second factor is the thermal broadening of these states, i.e. the existence of vibrationally excited states with a non-zero occupation probability.[18, 19] The corresponding reorganization energy after absorption or emission is frequently used to characterize disorder by extracting the width of a Gaussian fit to the data.[19-23] Alternatively, it is equally common to describe the experimentally observable disorder in absorption and emission spectra using an exponential Urbach tail whose temperature dependence provides some information about the relative importance of static and dynamic disorder.[21, 24] The debate is ongoing which effect is dominating, the energetic disorder of the actual density of states or the reorganization energy.[20, 21, 23, 25] However, both effects are detrimental for device performance via their impact on charge transport and recombination.[26-28] Thus, the experimental quantification of energetic disorder remains important independent of the exact origin of the disorder. There are several different methods to measure the disorder, the most prominent among them are optical methods where absorption or emission of a sample is recorded.[29-35] These are followed by electrical methods, where the voltage is varied to move the quasi-Fermi levels over the subband-gap density of states and thereby fill or empty the broadened density of states.[36-39] Given that observables such as the current or the capacitance depend on the carrier density inside a device under certain conditions, measurements such as charge extraction[36-38] or capacitance voltage[39] have been used to extract information on the band tails.

**Figure 1** shows how band tails impact photovoltaic performance and how typical values of the reported Urbach energies depend on the type of device and the mode of measurement. Figure 1a shows the simulated impact of the width of the Urbach tail on device performance using parameters typical for organic solar cells (see Table S1 in the Supporting Information). The histogram shown in Figure 1b provides evidence that the development of nonfullerene acceptors (NFAs) has not only improved power conversion efficiencies of organic solar cells but has also generally reduced the reported Urbach energies relative to the formerly predominant fullerene acceptors (FAs). A substantial number of reports of Urbach tails are now close to the thermal energy $kT$, which is highly significant for device performance from a theoretical and practical perspective.[40] Figure 1c shows a histogram of Urbach energies resolved for the general class of measurement technique. Here, we discriminate between optical techniques such as photothermal deflection spectroscopy (PDS) or Fourier-transform photocurrent spectroscopy (FTPS) that probe photon absorption and essentially measure the joint density of states and electrical techniques that use changes in applied voltage to scan the energy dependent density of states. Here, we note that optical techniques very frequently lead to fairly low Urbach energies around *kT* or slightly higher, while electrical techniques give values in a very broad range with a significant number of cases going up to 3*kT*. Thus, Figure 1c suggests that different methods may systematically lead to different Urbach energies, which would stipulate a closer look at the methods themselves. However, the assessment of Figure 1c is insofar incomplete as it provides only statistic evidence for a method specific difference, but does not provide data, where several techniques have been explored on the same samples.

Here, we provide such a study of different measurement techniques on the same samples and show how to overcome major issues with the measurement and interpretation of Urbach energies in NFA-based organic solar cells. We compare four different types of measurements done on two different types of organic solar cells. For the optical measurements, we performed PDS on the material films on glass and FTPS on full solar cells. These solar cells, we also characterized with electrical, voltage-dependent methods, where we chose admittance spectroscopy and extracted the Urbach energy from measurements in the dark and at open circuit. We find that Urbach energies indeed vary greatly depending on the methods used. To explain these discrepancies, we more closely investigate artefacts that are arising from capacitance-based measurements. We also highlight the importance of considering the different energy ranges accessed by the different methods. Thereby, we offer insights for a more conclusive characterization of energetic disorder in organic solar cells in the future.

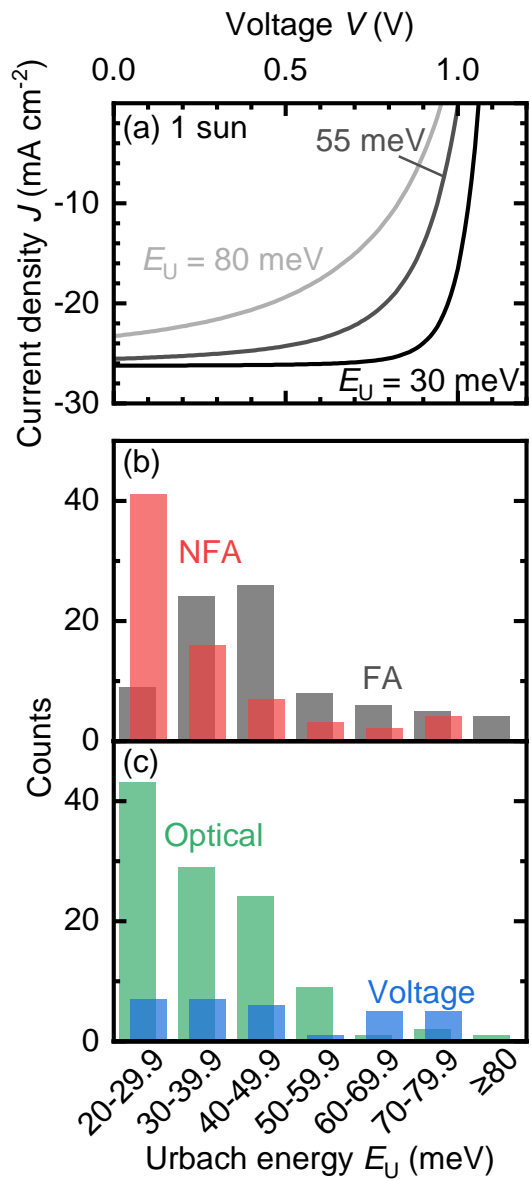

**Figure 1.** (a) Simulated current-density voltage curves of organic solar cells with increasing energetic disorder under solar illumination. The performance significantly deteriorates with increasing Urbach energy $E_U$. Still, Urbach energies reported in literature differ between (b) fullerene(FA)[7, 31, 35-37, 39, 41-73] and nonfullerene acceptors(NFA)[29-35, 38, 39, 47-50, 74-93] and especially between (c) optical[7, 29-35, 41, 42, 48-63, 74-90] and electrical, voltage-dependent[36-39, 45, 46, 69-72, 92, 93] measurement techniques highlighting inconsistencies in the characterization of energetic disorder in organic solar cells. Table S2 in the Supporting Information lists all materials, methods and references from this Figure.

## 2. Extracting the Urbach Energy from Admittance Spectroscopy

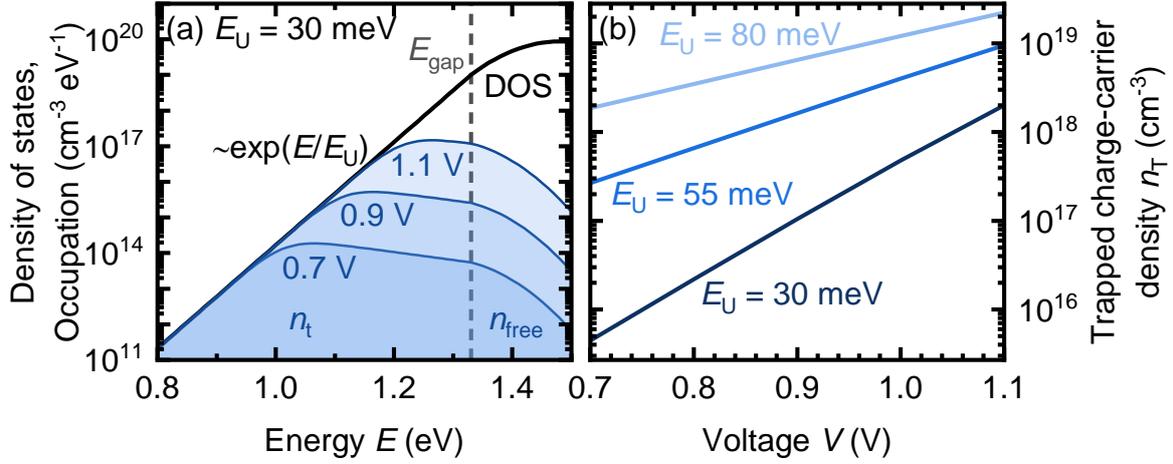

**Figure 2.** (a) Modelled density of states around the LUMO of the acceptor with exponential band tails. As the occupation maximum of the density of states is located at the quasi-Fermi level of electrons, the density $n_T$ of trapped electrons increases with quasi-Fermi level splitting. Consequently, $n_T$ increases with voltage in (b) for Urbach energies $E_U$ of 30, 55 and 80 meV. Low Urbach energies $E_U$ result in a more rapid increase since the density of tail states is steeper.

In a simplified case, the static disorder in the form of shallow defect states can be modelled by an exponential with the inverse slope $E_U$, the Urbach energy,[94] as shown in **Figure 2a**. When out of equilibrium, the density of states is occupied according to the Shockley-Read-Hall distribution by charge carriers with an occupancy of ½ at the quasi-Fermi level of the respective charge-carrier type which leads to an occupation maximum for an exponential density of tail states. With increasing applied voltage $V$, the quasi-Fermi level splitting increases and thereby, more charge carriers occupy the defect states as illustrated in Figure 2a for a device in the dark. The resulting density $n_T$ of trapped charge carriers therefore depends on the slope of the density of tail states. This dependence is further illustrated in Figure 2b which shows the $n_T$ as a function of voltage for three different Urbach energies $E_U$. The trapped charge-carrier density increases exponentially with voltage with a higher slope for a low Urbach energy since the corresponding density of tail states is steeper. Hence, the voltage dependence of the density of trapped charge-carriers $n_T$ can reflect the shape of the density of states. In a simplified case, this relation can be described by $n_T \sim \exp(qV/(2E_U))$ (see Section I of the Supporting Information for the derivation). As can be seen from Figure 2a, the area below the band edge that represents $n_T$, can be significantly larger than the one indicating the free carrier density $n_{\text{free}}$ in the band, allowing the approximation of the total charge-carrier density $n = n_T + n_{\text{free}} \approx n_T$. Hence, to extract the Urbach energy from electrical measurements, one can try to measure the charge-carrier density as a function of voltage.

Typical methods for finding $n$ are charge-extraction measurements and capacitance-voltage measurements from admittance spectroscopy. While the first measures the current that is extracted when switching a device from open circuit under illumination to short circuit in the dark, the latter uses the fact that a separation of charges is needed to create a capacitance inside a solar cell. A common approximation for this capacitance is

$$C_\mu = qd\frac{dn}{dV}, \tag{1}$$

where $C_\mu$ is the chemical capacitance per area of the charge carriers inside the bulk,[95, 96] in contrast to the electrode capacitance $C_\sigma$. Together, they form the total capacitance $C = C_\mu + C_\sigma$.[97] Note that here, as well as in charge-extraction measurements, charge-carrier densities are averaged over the active layer thickness. There is no spatial resolution. With this simplification and the previous assumptions on the total carrier density $n$, one can write

$$C_\mu \propto \frac{dn}{dV} \propto \frac{dn_T}{dV} \propto \frac{d}{dV}\left(\exp\left[\frac{qV}{2E_U}\right]\right) \propto \exp\left[\frac{qV}{2E_U}\right]. \tag{2}$$

Hence, from the logarithmic slope of the chemical capacitance $C_\mu$, the Urbach energy $E_U$ can be extracted in this model. In literature, the chemical capacitance most commonly is further integrated to get the charge-carrier density $n$ and interpret it in terms of recombination mechanisms.[34, 39, 97-101] While the correlation between $n$ and the recombination rate is relatively intuitive, there are multiple approaches and some discussion on how to proceed with this integration and how to estimate the chemical capacitance.[97, 100, 102] A more detailed analysis of the issues of extracting the Urbach energy from the charge-carrier density can be found in the Supporting Information, Section II. For these reasons, we herein refrain from using further calculation steps during integration and directly use the slope of the chemical capacitance for the Urbach-energy estimation.

To validate this approach and find the most accurate way to calculate the chemical capacitance from admittance data, we modelled a generic organic solar cell using the drift-diffusion simulation software SCAPS.[103, 104] For simulations under illumination, we further used the software ASA[105, 106] and optical data from Reference [107] to create the spatially resolved generation rate. Further details on the model and the simulation parameters can be found in the Supporting Information, Section III. Since the extraction of the chemical capacitance from admittance data is not obvious, we first performed simulations to find the best estimation for $C_\mu$. These simulations show that the chemical capacitance can be best replicated by subtracting the total capacitance at high frequency and reverse bias in the dark, the geometric capacitance $C_{geo}$, from the total capacitance at low frequencies. Section IV in the Supporting Information describes in more detail how we calculate the chemical capacitance in this work.

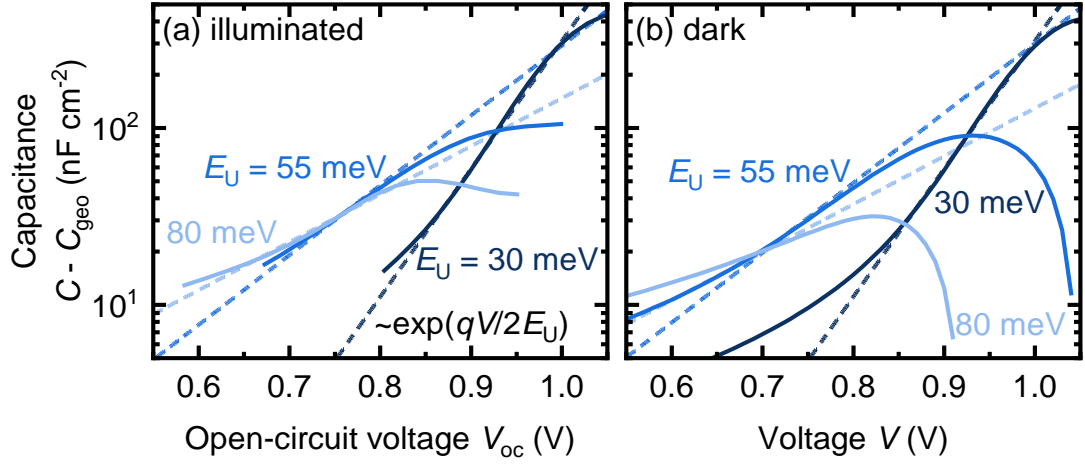

**Figure 3.** Capacitance $C$ - $C_{geo}$ extracted from simulated admittance spectroscopy measurements as a function of voltage for Urbach energies $E_U$ of 30, 55 and 80 meV. The dashed lines indicate an exponential with the corresponding Urbach energy. For both operating conditions, (a) under illumination at open circuit and (b) in the dark, the quasi-Fermi level splitting can be approximated by the applied voltage and thereby the chemical capacitance can indicate the respective Urbach energy.

Now being able to estimate the chemical capacitance from the simulated admittance data, we test the method of extracting the Urbach energy from the $C_\mu \sim V$-relation. For this purpose, **Figure 3** shows the chemical capacitance $C_\mu$ of three organic solar cells with different tail slopes $E_U$ of 30, 55 and 80 meV that were calculated from admittance spectroscopy simulations (a) under illumination at open circuit and (b) in the dark using the approximation $C_\mu \approx C - C_{geo}$. The dashed lines represent the exponential increase that can ideally be expected for the respective Urbach energies. In fact, under both conditions where the quasi-Fermi level splitting can be approximated by the applied voltage, in the dark and at open circuit, the capacitance $C$ - $C_{geo}$ follows the exponential increase to some extent. This observation shows that the chemical capacitance is sensitive to the shape of the density of defect states and can be used for an estimation of the Urbach energy. Still, this introduction to the rational behind the characterization of the disorder with electrical methods already hints towards a different level of complexity compared to the optical methods. Here, the absorptance is directly related to the density of defect states that allow the absorption of photons with energies below the energy gap. Therefore, less approximations go into the analysis of optical experimental data that we present in the following section.

## 3 Experimental results

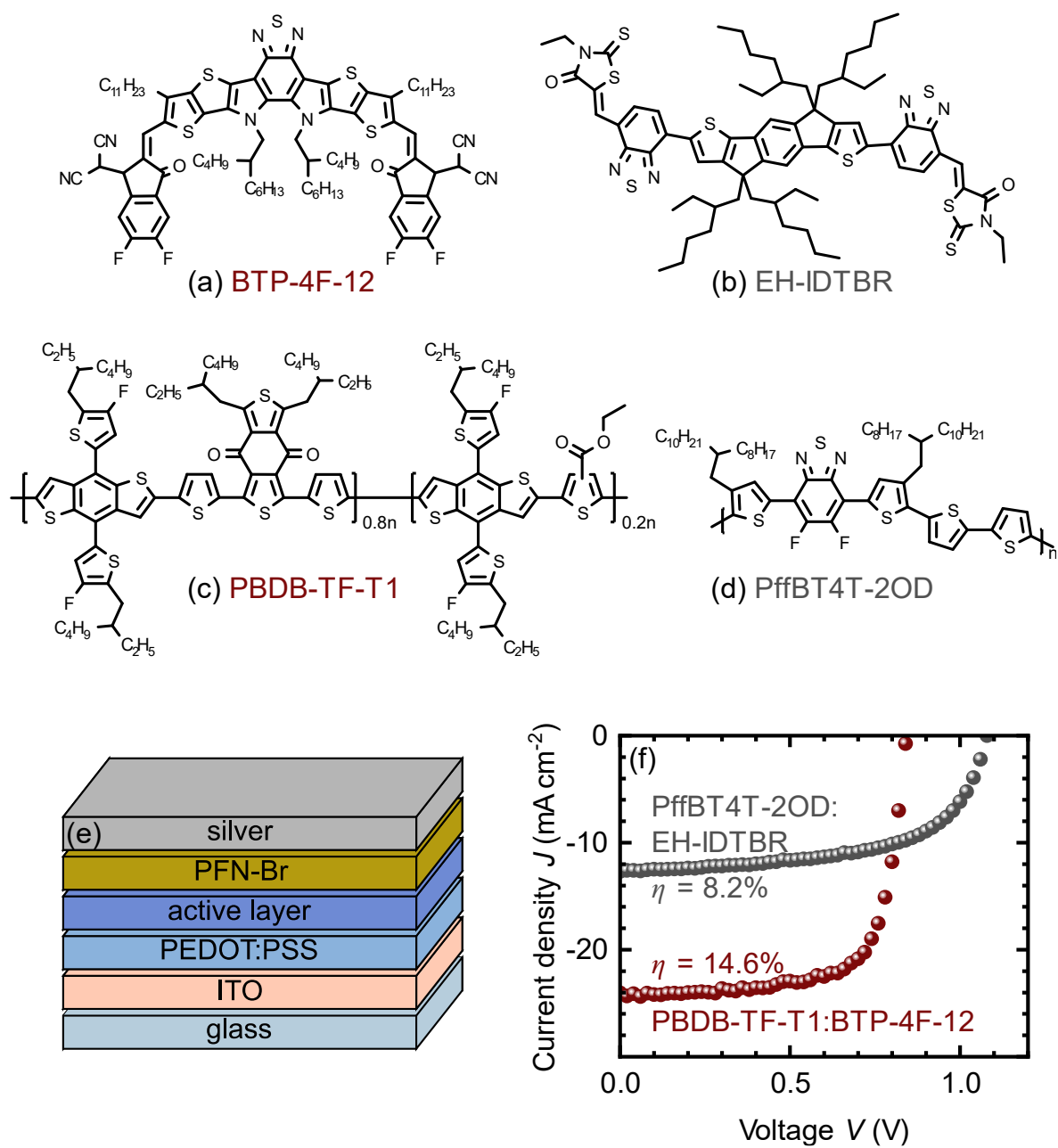

**Figure 4.** (a,b) Nonfullerene acceptors and (c,d) polymer donors used for solar cell production in the (e) device architecture glass/ITO/PEDOT:PSS/active layer/PFN-Br/Ag. (f) Current density $J$ as a function of voltage $V$ of the two organic solar cells incorporating the bulk heterojunctions PffBT4T-2OD:EH-IDTBR and PBDB-TF-T1:BTP-4F-12 that are characterized in this work. Urbach energies reported in literature for the same or similar blends imply different severity of energetic disorder between the two material systems.

To study the difference between optical and voltage-dependent measurements of the Urbach energy $E_U$, we characterized two organic solar cells with identical cell architecture but different active layer material systems. More specifically, we used the nonfullerene acceptors (2,2´-((2Z,2´Z)-((12,13-bis(2-butyloctyl)-3,9-diundecyl-12,13-dihydro-[1,2,5]thiadiazolo[3,4-e]thieno[2,"3´´:4',5´]thieno[2´,3´:4,5]pyrrolo[3,2-g]thieno[2´,3´:4,5]thieno[3,2-b]indole-2,10-diyl)bis(methanylylidene))bis(5,6-difluoro-3-oxo-2,3-dihydro-1H-indene-2,1-diylidene))dimalononitrile) (BTP-4F-12, **Figure 4a**) and (Z)-5-{[5-(15-{5-[(Z)-(3-Ethyl-4-oxo-2-thioxo-1,3-thiazolidin-5-ylidene)methyl]-8thia-7.9-diazabicyclo[4.3.0]nona-1(9),2,4,6-tetraen-2-yl}-9,9,18,18-tetrakis(2-ethylhexyl)-5.14-dithiapentacyclo[10.6.0.0$^{3,10}$.0$^{4,8}$.0$^{13,17}$]octadeca-1(12),2,4(8),6,10,13(17),15-heptaen-6-yl)-8-thia-7.9diazabicyclo[4.3.0]nona-1(9),2,4,6-tetraen-2-yl]methylidene}-3-ethyl-2-thioxo-1,3-thiazolidin-4one (EH-IDTBR, Figure 4b) and the polymer donors Poly[(2,6-(4,8-bis(5-(2-ethylhexyl-3-fluoro)thiophen-2-yl)-benzo[1,2-b:4,5-b']dithiophene))-alt-(5,5-(1',3'-di-2-thienyl-5',7'-bis(2-ethylhexyl)benzo[1',2'-c:4',5'-c']dithiophene-4,8-dione)]-ran-poly[(2,6-(4,8-bis(5-(2-ethylhexyl)thiophen-2-yl)-benzo[1,2-b:4,5-b']dithiophene))-alt-(2,2-ethyl-3(or4)-carboxylate-thiophene)] (PBDB-TF-T1, Figure 4c) and Poly[(5,6-difluoro-2,1,3-benzothiadiazol-4,7-diyl)-alt-(3,3'''-di(2-octyldodecyl)-2,2';5',2'';5'',2'''-quaterthiophen-5,5'''-diyl)] (PffBT4T-2OD, Figure 4d). The energetic disorder of the high efficiency material system PBDB-TF-T1:BTP-4F-12 to the best of our knowledge has not been investigated so far. Similar materials have been found to exhibit a low Urbach energy.[38, 80, 93] As a contrast, we chose PffBT4T-2OD:EH-IDTBR as an active layer blend for which a relatively high Urbach energy of 76 meV has been reported.[38] Both solar cells were fabricated on a glass substrate with an indium tin oxide (ITO) anode, and Poly(3,4-ethylenedioxythiophene) polystyrene sulfonate (PEDOT:PSS) as the hole transport layer. On top of the active layer, Poly(9,9-bis(3'-(N,N-dimethyl)-N-ethylammoinium-propyl-2,7-fluorene)-alt-2,7-(9,9-dioctylfluorene))dibromide (PFN-Br) creates the electron transport layer and a silver cathode is used. While a schematic of the entire stack is shown in Figure 4e, further information on the fabrication process of the organic solar cells can be found in the Supporting Information,

Section V. Details on the measurement specifications during characterization are listed in Section VI of the Supporting Information. Figure 4f shows the *JV*-characteristics of the resulting devices. The cell with PffBT4T-2OD:EH-IDTBR exhibits a higher open-circuit voltage due to its higher band gap but significantly less photocurrent than the device with PBDB-TF-T1:BTP-4F-12. However, it is beyond the scope of this work to analyze the various loss mechanisms that set apart the two solar cell systems. Instead, we want to focus on the recombination via shallow defect states.

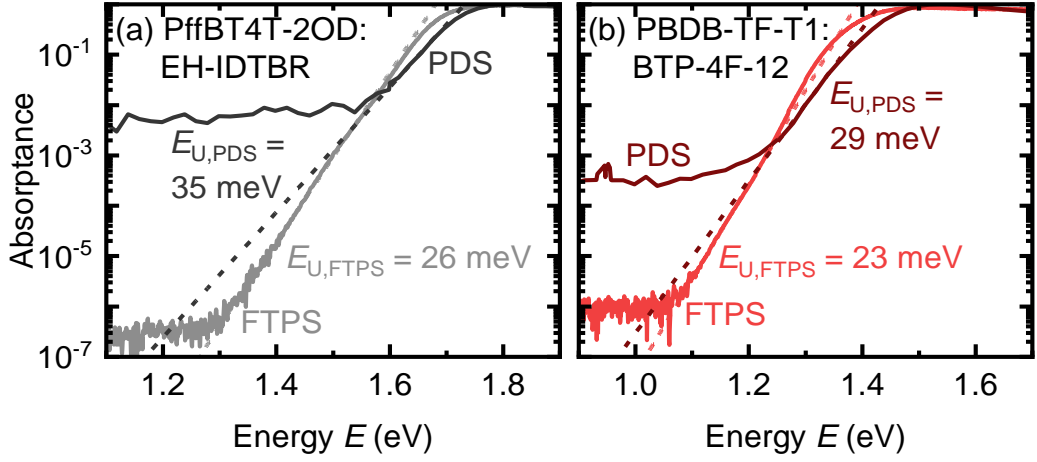

**Figure 5.** Normalized signal of Fourier-transform photocurrent spectroscopy (FTPS) in the lighter colors and photothermal deflection spectroscopy (PDS) in the darker colors as a function of energy $E$ for organic solar cells based on (a) PffBT4T-2OD:EH-IDTBR and (b) PBDB-TF-T1:BTP-4F-12. The dashed lines represent fits to the exponential regime of the experimental data with the slope of $1/E_U$. Both material systems exhibit Urbach energies $E_U$ close to the thermal energy.

For this purpose, we characterized the absorption properties of the devices with Fourier transform photocurrent spectroscopy (FTPS) and of the active layer films with photothermal deflection spectroscopy (PDS). **Figure 5** shows the normalized signal of the optical measurements on the material systems (a) PffBT4T-2OD:EH-IDTBR and (b) PBDB-TF-T1:BTP-4F-12. For every measurement, we fitted the exponential regime (dashed lines) to extract the Urbach energy $E_U$ from the slope $1/E_U$. Under FTPS, both solar cells show a very steep increase in absorptance which results in an Urbach energy $E_{U,FTPS}$ of around 26 meV for PffBT4T-2OD:EH-IDTBR and 23 meV for PBDB-TF-T1:BTP-4F-12. While the FTPS measurements show a dynamic range of around six orders of magnitude, PDS has a lower range between two and three orders of magnitude. Therefore, the exponential part of the signal could be affected by the saturation at a higher energy resulting in higher Urbach energies $E_{U,PDS}$ of 35 meV for PffBT4T-2OD:EH-IDTBR and 29 meV for PBDB-TF-T1:BTP-4F-12. Also, the films on glass that were measured for PDS lack absorption from back reflection on the silver cathode that is included in the FTPS measurements causing further discrepancy between the two methods. Still, both optical measurements yield relatively low Urbach energies indicating low energetic disorder.

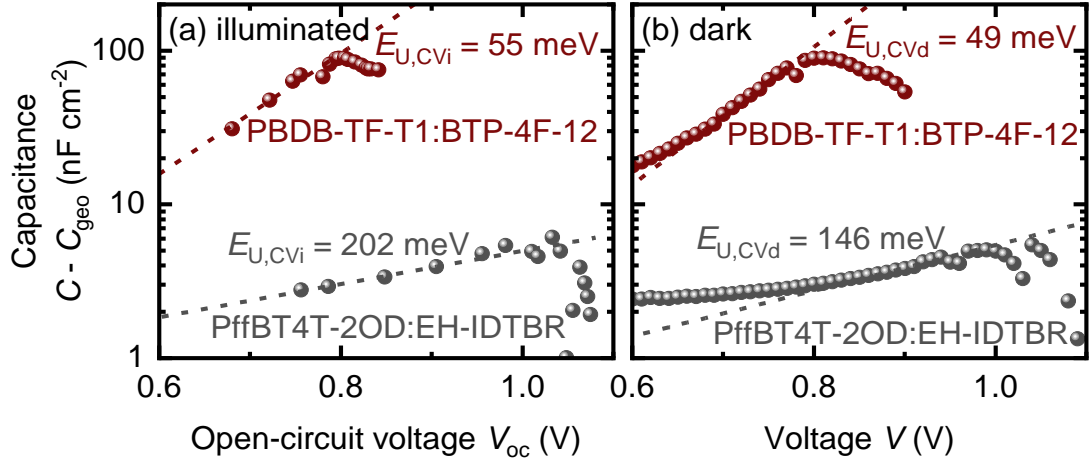

**Figure 6.** Capacitance $C - C_{geo}$ estimated from admittance spectroscopy measurements (a) under illumination at open circuit and (b) in the dark on organic solar cells with an active layer consisting of PffBT4T-2OD:EH-IDTBR or PBDB-TF-T1:BTP-4F-12. The dashed lines represent exponential fits to the experimental data with a slope $1/(2E_{U,CV})$. The Urbach energy $E_{U,CV}$ extracted from the fits for PBDB-TF-T1:BTP-4F-12 is about twice as high as for the optical measurements whereas the values for PffBT4T-2OD:EH-IDTBR are unrealistically high.

For the voltage-dependent method, we performed admittance spectroscopy on the same solar cells that were characterized by FTPS. From the experimental admittance data, we calculated the capacitance $C - C_{geo}$ as discussed previously as an estimate for the chemical capacitance $C_\mu$. **Figure 6a** shows the resulting capacitance as a function of open-circuit voltage $V_{oc}$ for both solar cells based on PffBT4T-2OD:EH-IDTBR and PBDB-TF-T1:BTP-4F-12. At low voltages, the capacitance increases before falling at higher voltages. When fitting an exponential function to the increase, one can extract an Urbach energy $E_{U,CVi} = 55$ meV for PBDB-TF-T1:BTP-4F-12 from the data and $E_{U,CVi} = 202$ meV for PffBT4T-2OD:EH-IDTBR. The capacitance $C - C_{geo}$ in the dark in Figure 6b behaves similarly resulting in $E_{U,CVd} = 49$ meV and $E_{U,CVd} = 146$ meV for PBDB-TF-T1:BTP-4F-12 and PffBT4T-2OD:EH-IDTBR, respectively. Therefore, the tail slope extracted from voltage-dependent admittance measurements is more than twice as high for PBDB-TF-T1:BTP-4F-12 as the one from optical measurements. This difference appears to be well in line with our observation on Urbach energies reported in literature. The values extracted from fits of the capacitance of the PffBT4T-2OD:EH-IDTBR solar cell, however, are higher than any typically reported in literature.

This difference between the methods cannot be explained by including the reorganization energy into the interpretation. While optical techniques measure both, the reorganization energy and static disorder,[20, 21, 23] electrical methods only measure the filling of the lowest excited state with electrons and the highest ground state with holes. So, as there is no charge transfer needed between the states, the reorganization energy is not reflected in the capacitance data. Therefore, this consideration would even lead to higher disorder measured by optical characterization contrary to what we observed in our experiments and in literature. To understand these inconsistencies between measurement techniques, we further used drift-diffusion simulations to study possible origins leading to higher Urbach energies from voltage-dependent measurements. As the drastic difference in Urbach energy $E_{U,CV}$ between the two material systems suggests that different effects may occur in the two devices, we will in the following look at the solar cells based on PBDB-TF-T1:BTP-4F-12 and PffBT4T-2OD:EH-IDTBR separately.

## 4 Discussion

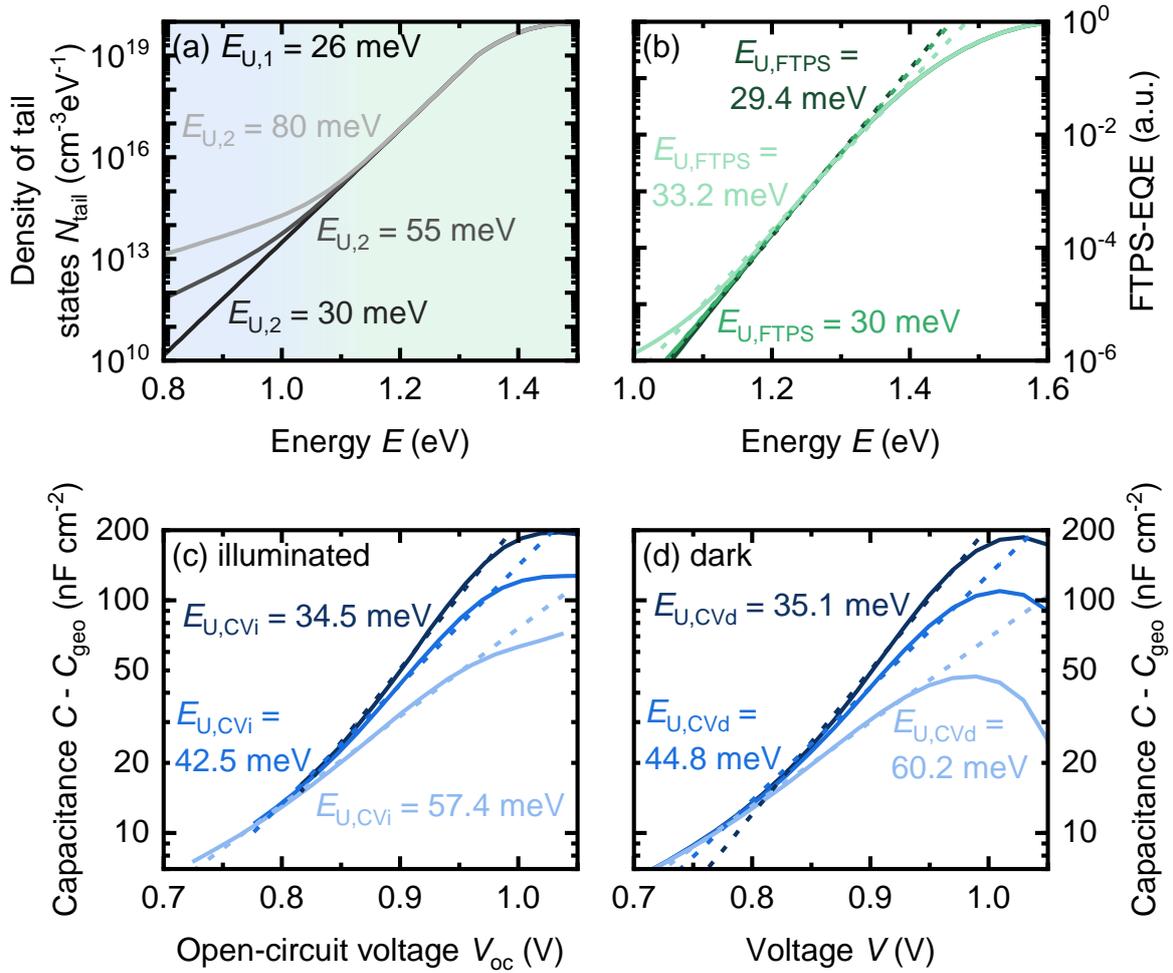

**Figure 7.** (a) Total density of states modelled with a combination of a steep Urbach tail with $E_{U,1} = 26$ meV that dominates near the band edge and a tail with higher Urbach energies $E_{U,2}$ of 30, 55 or 80 meV that dominates at lower energies $E$. (b) FTPS-signal calculated from the density of states with two exponential tails. (c) Corresponding simulated capacitance $C - C_{geo}$ under open-circuit conditions and (d) in the dark. The Urbach energies extracted from the dashed fits of the capacitance measurements show more sensitivity to the slope of the deeper tail than for the optical measurements.

We first focus on the effect observed in the PBDB-TF-T1:BTP-4F-12 cells. For this purpose, we modelled a solar cell with a density of tail states consisting of two tails with different slopes and trap density of states. **Figure 7a** shows the resulting density of tail states for a steep Urbach tail with $E_{U,1} = 26$ meV that dominates close to the band edge and a more shallow tail with varying Urbach energy $E_{U,2}$ of 30, 55 and 80 meV that dominates further in the band. The FTPS signal that can be calculated from this density of states is displayed in Figure 7b. In the range that can be resolved by the measurement, there is only little influence by the deep tail. Therefore, the Urbach energy $E_{U,FTPS}$ that could be extracted from these optical measurements only ranges from 29 to 33 meV. In contrast, the admittance measurements in Figure 7c and Figure 7d are more sensitive to the variation of the deep Urbach tail and result in significantly different fits. Therefore, at the voltages applied to the solar cell, these deep tails are still filled with increasing voltage. So, with the filling according to the quasi-Fermi levels, an energy range of the density of states is probed that can be below the resolution of FTPS measurements. Thereby, the voltage-dependent measurements can in fact show features of deeper traps in the device. Hence, the difference in Urbach energy between optical and voltage-dependent measurements as observed for PBDB-TF-T1:BTP-4F-12 can originate in probing a density of trap states at different energy ranges that is not monoexponential.

Yet, for this effect to evoke Urbach energies $E_{U,CV}$ as high as observed for PffBT4T-2OD:EH-IDTBR, the slope of the density of deep defect states would need to be extremely low. Therefore, we need to further study the chemical capacitance of this material system and explore the possibility that it does not actually reflect the density of states. As seen in Figure 4f, the solar cell based on PffBT4T-2OD:EH-IDTBR only shows a power-conversion efficiency of around 8%. The material system is known to have poor charge-carrier mobilities[108] and due to different energy levels compared to PBDB-TF-T1:BTP-4F-12, the injection barriers might differ. In addition, even though the density of states seems to be steep close to the band edge, recombination via these tail states can still be high due to high capture rates. Even the increased active layer thickness which is around 150 nm for this solar cell can lead to transport issues and space-charge effects due to less uniform charge-carrier generation. The effect of the injection barriers, the capture rates and the active layer thickness on the capacitance-voltage curves predicted by simulations is shown in Figure S6 in the Supporting Information. In the following we will focus only on the charge-carrier mobility as an example.

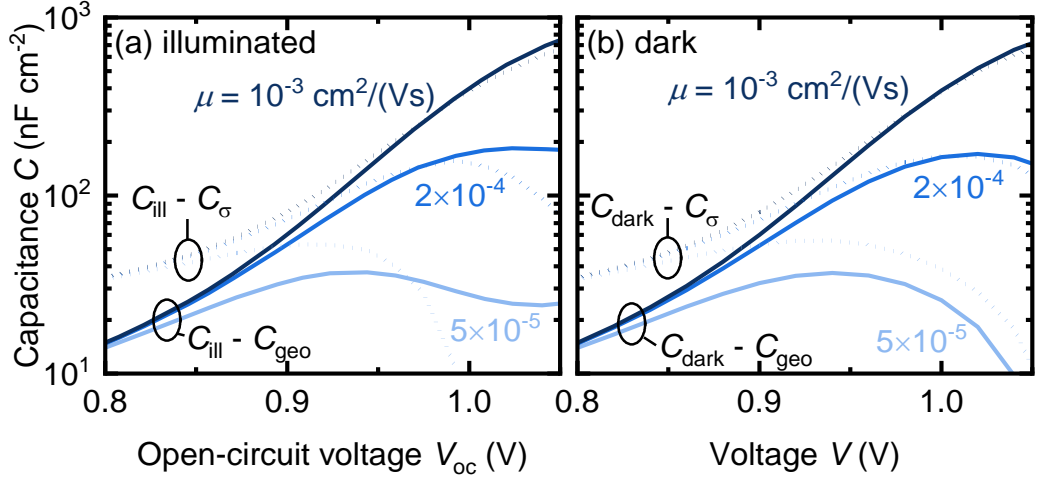

**Figure 8.** Simulated capacitance (a) at open circuit under varying illumination intensity and (b) in the dark for different applied voltages $V$ and multiple charge-carrier mobilities $\mu$. The solid lines represent the difference between the total capacitance and the geometric capacitance $C_{\text{geo}}$, which is an estimate of the chemical capacitance that can be determined experimentally. The curves flatten for lower mobilities. The dotted lines are the difference between the total capacitance and the electrode capacitance $C_\sigma$, which is the best approximation for $C_\mu$ but can only be extracted from simulations. The issue of flat $CV$-curves does not originate in the assumption $C_\sigma \approx C_{\text{geo}}$, since both approaches show a similar behavior at high voltages with decreasing charge-carrier mobility.

**Figure 8** shows the simulation results for an Urbach energy $E_{\text{U,CV}}$ of 30 meV but different charge-carrier mobilities $\mu$ to illustrate how limited transport can cause the experimental trends observed in Figure 6. Once more, it shows the chemical capacitance at (a) open circuit and (b) in the dark estimated by the difference between the total capacitance and the geometric capacitance $C_{\text{geo}}$ with the solid lines. It appears that under both working conditions, the curves are flattening with decreasing mobilities. Therefore, at low charge-carrier mobilities, the Urbach energy $E_{\text{U,CV}}$ calculated from the slope of the graphs would be an overestimation. There are several assumptions underlying this estimation. One of them is the calculation of the chemical capacitance as the difference $C - C_{\text{geo}}$ of the total capacitance and the constant geometric capacitance $C_{\text{geo}}$ as discussed in the Supporting Information in Section IV. It assumes that the charge on the electrodes that contributes to the geometric capacitance $C_{\text{geo}}$ is constant over the entire voltage range. In reality, the electrode capacitance can be voltage dependent. Therefore, for the true chemical capacitance, one needs to subtract the voltage-dependent electrode capacitance $C_\sigma$. Even though $C_\sigma$ is not experimentally accessible, we can extract it from the drift-diffusion simulations. Consequently, Figure 8 shows $C - C_\sigma$ with the dashed lines for comparison. It reveals the same trend with mobility as the experimentally accessible $C - C_{geo}$. So, the flattening of the *CV*-curves does not originate in capacitive but rather resistive effects.

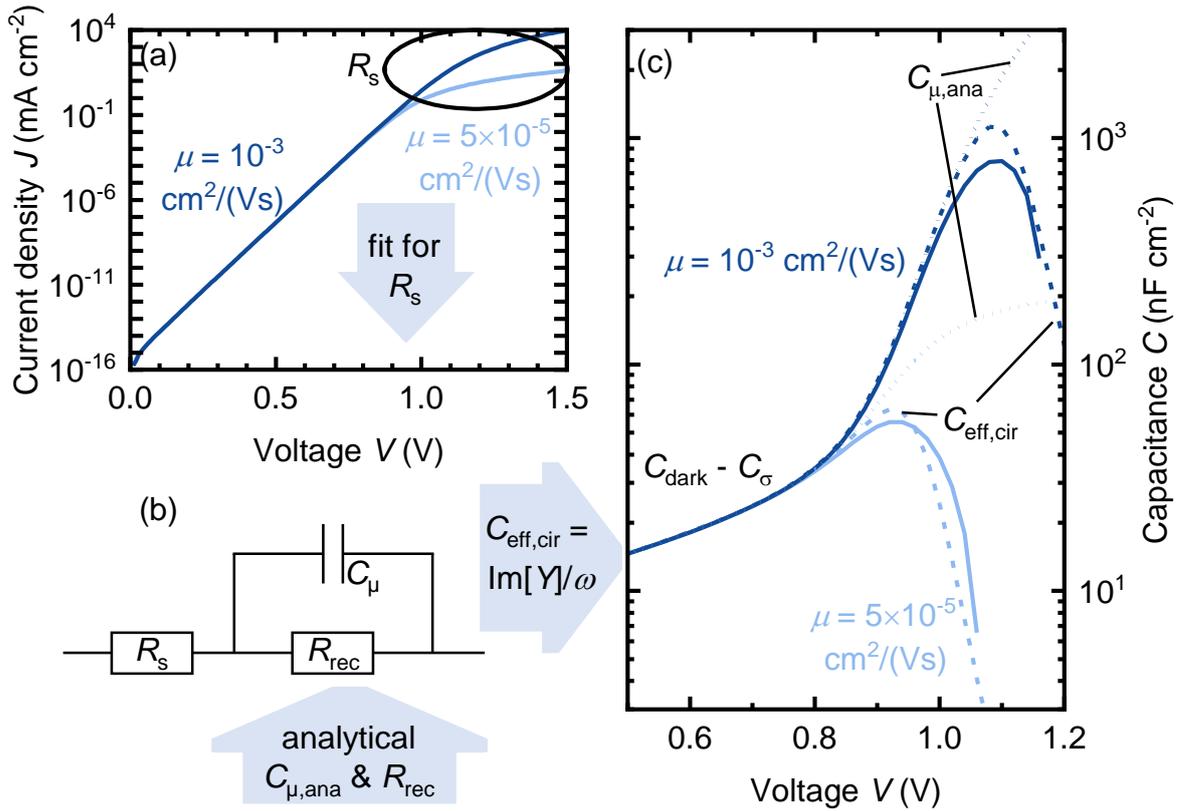

**Figure 9.** (a) Dark *JV*-characteristic of organic solar cells with high and low charge-carrier mobilities $\mu$ and no external resistance. The high-voltage regime indicates an internal series resistance $R_s$. (b) Equivalent circuit model extended by the internal series resistance $R_s$ which can be estimated from fitting the dark *JV*-curve. The chemical capacitance $C_{\mu,\text{ana}}$ and the recombination resistance $R_{\text{rec}}$ can be calculated from steady state quantities using analytical expressions. (c) Dark capacitance-voltage characteristics for high and low mobilities $\mu$. The effective capacitance $C_{\text{eff,cir}}$ with the dashed lines calculated from the equivalent circuit fits well with the chemical capacitance $C - C_\sigma$ given by the drift-diffusion simulations. The constant increase in the dotted analytical chemical capacitance $C_{\mu,\text{ana}}$ proves that the internal series resistance $R_s$ is needed to explain the drop in the simulated capacitance.

**Figure 9** illustrates how a series resistance $R_s$ inside the active layer causes the flattening of the capacitance-voltage curves. In Figure 9a, the dark current density is plotted for an organic solar cell with high mobilities and one with low mobilities. Even though external resistances are neglected in the simulations, the current density increases less rapidly at high voltages, typical for a series resistance. This behavior indicates that the limited mobility causes an internal series resistance $R_s$ that increases with decreasing charge-carrier mobility. Therefore, the simple RC-circuit that was used so far for the determination of the total capacitance as the imaginary part of the admittance $Y$ must be extended by a series resistance. The resulting equivalent circuit is shown in Figure 9b. While the series resistance $R_s$ can be roughly estimated by fitting the dark $JV$-curves with the diode equation, the two other circuit elements, the chemical capacitance $C_\mu$ and the recombination resistance $R_{rec}$, can also be obtained from steady-state quantities given by the simulations. An estimation of all circuit elements allows us to calculate the effective capacitance $C_{eff,cir}$ that we would get if the equivalent circuit in Figure 9b was true, but we still assumed $C_{eff,cir} = \text{Im}(Y)/\omega$, where $\omega$ is the angular frequency. For further details on the calculation of $C_{eff,cir}$, see Section VII in the Supporting Information. Figure 9c shows this effective capacitance $C_{eff,cir}$ for the high and low mobility $\mu$ alongside the capacitance $C - C_\sigma$ that was discussed in Figure 8b. The latter is given by the imaginary part of the admittance simulated under alternating current. In fact, the effective capacitance $C_{eff,cir}$ based on the circuit in Figure 9b replicates the decreasing capacitance $C - C_\sigma$ very well. The chemical capacitance $C_{\mu,ana}$ calculated analytically from steady state quantities shows that the chemical capacitance itself is not dropping. Only the inclusion of the series resistance $R_s$ allows for a drop in effective capacitance. Therefore, all quantities that hinder transport and thereby cause an internal series resistance $R_s$ will lead to an early decrease in the effective capacitance and therefore an overestimation of the Urbach energy $E_{U,CV}$. In contrast, transport issues are not relevant for the PDS measurement and much less significant in FTPS measurements due to the stronger internal electric field under the operation at short circuit. Thus, the high susceptibility of admittance measurements to bad electronic properties can lead to the discrepancies in the reported Urbach energy compared to the less sensitive optical measurements.

## 5  Conclusions

The characterization of energetic disorder plays an important role in the search for material systems that enable high efficiency organic solar cells. Optical methods are often used for this purpose where electrons are excited from or into defect states by incident photons. For instance, photothermal deflection spectroscopy (PDS),[29-31] Fourier transform photocurrent spectroscopy (FTPS)[32, 33] or highly sensitive external quantum efficiency (EQE) measurements[34, 35] are frequently applied in literature. But apart from these absorption-based methods, the density of defect states can also be probed by varying the quasi-Fermi level splitting with an applied voltage in, for example, charge extraction[36-38] or admittance measurements.[39] We found that both in literature and in our experiments these electrical, voltage-dependent measurements yield overall higher Urbach energies than the techniques based on optical excitation. We conducted optical FTPS and PDS measurements and electrical admittance spectroscopy under illumination and in the dark on two nonfullerene acceptor-based material systems. In all cases, the Urbach energy extracted by the voltage-dependent methods was at least twice as high as their optical counterparts.

In our analysis, we show that care must be taken when analyzing experimental data as different effects can be mistakenly interpreted as features of energetic disorder. Even for purely optical data, we have observed that a low dynamic range in PDS measurements can lead to a lower slope at the band edge than in FTPS on the same material. Voltage-dependent admittance measurements have proven to be even more delicate as the analysis in terms of energetic disorder is based on a high number of assumptions. Our experimental and simulation results on the extreme case of a solar cell based on PffBT4T-2OD:EH-IDTBR have demonstrated how bad transport properties can lead to an internal series resistance overlaying the exponential regime of the capacitance-voltage measurements. Thereby, the discrepancy in the Urbach energy between optical and voltage-dependent measurements can originate in an overestimation in voltage-dependent measurements due to bad electronic properties.

Knowing the limitations of the characterization techniques, we have also highlighted the potential that both optical and electrical methods correctly reflect the subband-gap density of states even though they yield different values for the Urbach energy. Moving away from a strictly monoexponential band tail, we have shown that the quasi-Fermi level splitting that is typical for voltage-dependent measurements probes energy ranges of the density of states where the signal of the optical measurements is below its resolution. Therefore, different approaches may detect different features of a density of defect states.

So, we recommend the combination of different characterization techniques to not only be able to minimize the chance of unknowingly running into the limitations of a method but also to maximize the information gain on the energetic disorder in organic solar cells.

**Acknowledgements**

The authors acknowledge funding from the Helmholtz Association. S.R. acknowledges the German Research Foundation (DFG) for support through a Walter-Benjamin fellowship (Project No. 462572437).

Supporting Information

**Comparing Methods of Characterizing Energetic Disorder in Organic Solar Cells**

*Paula Hartnagel, Sandheep Ravishankar, Benjamin Klingebiel, Oliver Thimm and Thomas Kirchartz\**

I.  **Relation between the Urbach energy and the trapped carrier density**

All evaluation of the electrically acquired data in terms of the Urbach energy is based on the relation between the density of trapped carriers and the voltage. Here, we derive this relation on the example of the traps in the conduction-band tail (CBT). As we assume equal tails for conduction and valence-band tails in this work, the derivation of the trapped hole density follows analogously. For exponential tail states, the density of conduction-band tails

$$N_{\text{cbt}}(E) \propto \exp\left(\frac{E}{E_U}\right), \tag{S1}$$

as described in Figure 2a in the main paper. These trap states are occupied with a certain occupation probability $f(E)$. The integral over all occupied states then gives the density of trapped electrons

$$n_t \propto \int_{E_V}^{E_C} f(E) N_{\text{cbt}}(E) dE, \tag{S2}$$

where $E_V$ is the valence-band edge and $E_C$ the conduction-band edge. As the density of states is filled mostly until the quasi-Fermi level $E_{\text{qFn}}$ of the electrons, the occupation probability can be approximated by a step function and the integral becomes

$$n_t \propto \int_{E_V}^{E_{\text{qFn}}} N_{\text{cbt}}(E) dE. \tag{S3}$$

Solving this integral leads us to

$$n_t \propto \exp\left(\frac{E_{\text{qFn}}}{E_U}\right) - \exp\left(\frac{E_V}{E_U}\right). \tag{S4}$$

The term with the valence-band energy is negligibly small compared to the first term. Also, as the quasi-Fermi level splitting is evoked by an applied voltage and we consider symmetric devices, it can be further assumed that

$$n_t \propto \exp\left(\frac{qV}{2E_U}\right). \tag{S5}$$

This relation allows us to extract information on the density of trap states in the form of the Urbach energy from carrier-density data.

## II. Calculating the Urbach Energy from Charge-Carrier Densities

The charge-carrier density is frequently used to characterize recombination in organic solar cells.[1-7] One possible way to measure an apparent charge-carrier density is admittance spectroscopy. The method is based on the approximation of the chemical capacitance $C_\mu \sim dn/dV$, which implies that the charge carrier density $n$ can be extracted from the integral

$$n(V) = \frac{1}{qAd} \int_{-\infty}^{V} C_\mu(V) dV. \tag{S6}$$

Since an integration to infinitely low voltages is experimentally impossible, the lower integral boundary is set to a voltage $V_{sat}$, where the capacitance is saturated. This voltage differs amongst publications between -5 V,[7] -4 V,[3] -3 V,[2, 6] 0 V[1] or even higher.[5] In addition, Equation (S6) is altered to account for the charge-carrier density $n_{sat}$ that remains inside the solar cell even at high reverse bias. The result is

$$n(V) = n_{sat} + \frac{1}{qAd} \int_{V_{sat}}^{V} C_\mu(V) dV. \tag{S7}$$

However, there are several different definitions for $n_{sat}$ in literature.[5, 7] Vollbrecht et al even stated that the use of $n_{sat}$ results in an overestimation of the charge-carrier density since it includes charge carriers that are also covered by the integral.[8] This additional, slightly arbitrary constant can cause changes in the slope of the charge-carrier density in the semilogarithmic plot and therefore alter the Urbach energy $E_U$ extracted. For that reason, we herein refrain from the use of the saturation-charge carrier density $n_{sat}$ and calculate

$$n(V) = \frac{1}{qAd} \int_{V_{sat}}^{V} C_\mu(V) dV \tag{S8}$$

with $V_{sat} = -3$ V. Still, simply the lack of an accurate calculation technique for the charge-carrier density makes a correct extraction of the Urbach energy $E_U$ more unlikely.

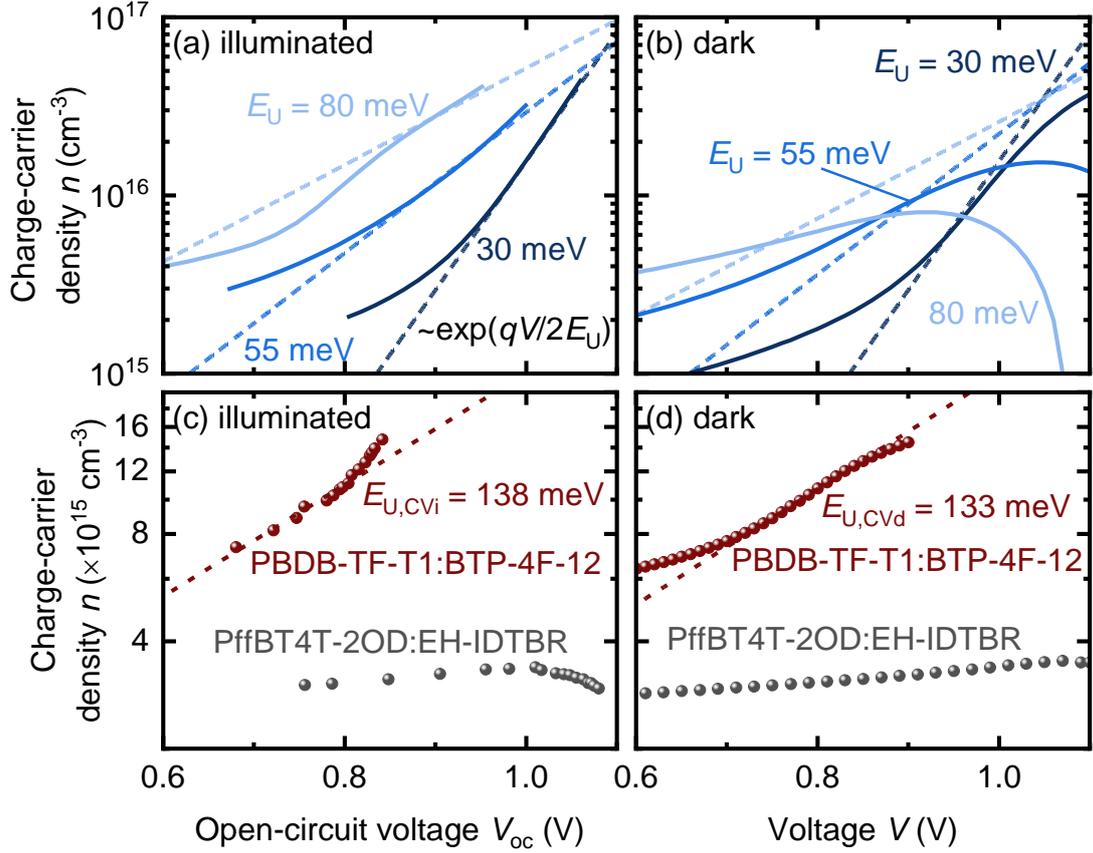

**Figure S1.** Charge-carrier density $n$ calculated from the integration of the (a, b) simulated and (c, d) experimental capacitance $C - C_{geo}$ as a function of applied voltage. (a) Under illumination, the lines simulated with different Urbach energies $E_U$ approach the correct slope whereas (b) in the dark, $n$ is sensitive to the Urbach energy but does not yield sensible slopes. In the experiment, both (c) illuminated and (d) dark charge-carrier density give values more than twice as high as for the chemical capacitance in Figure 6 of the main paper in the case of PBDB-TF-T1:BTP-4F-12.

The simulated charge-carrier density $n$ calculated with Equation (S8) can be seen in Figure S1a and S1b. While the charge-carrier density under illumination fits the dashed line of the correct slope similarly well as the chemical capacitance in Figure 3a, the dark charge-carrier density does not exhibit the correct slope at any point. One possible reason for this discrepancy between illuminated and dark measurements lies once again in the calculation technique. The charge-carrier density in the dark in Figure S1b shows the accumulated integral over the chemical capacitance in Figure 3b calculated with Equation (S8). In contrast, $n$ in Figure S1a is not the integral over $C_\mu$ in Figure 3a, but each point represents the integral from $V_{sat}$ to $V_{oc}$ of $C_\mu$ at a certain light intensity. So, opposed to the estimation of $E_U$ from $C_\mu$, where both measurements in the dark and under illumination yield similar results, for the charge-carrier density more care has to be taken when choosing the working point.

Figure S1c and S1d show the experimental results of the charge-carrier density calculated from $C_\mu$ as a function of voltage. Both in the dark and under illumination, the slope of the charge-carrier density for PBDB-TF-T1:BTP-4F-12 is about twice as high as the one from $C_\mu$ in Figure 6. This observation may be counterintuitive at first. When integrating an exponential function such as $C_\mu = C_{sc} \exp(qV/2E_U)$, where $C_{sc}$ is the capacitance at short circuit, the slope of the exponential should remain the same. In the experiment, however, the chemical capacitance saturates to a nonzero value and therefore has the shape $C_\mu = C_{sc} \exp(qV/2E_U) + \text{const.}$. When integrating $C_\mu$ with this additive term, the slope of the charge-carrier density is altered. Hence, the additive constant makes an estimation of $E_U$ from $n$ increasingly difficult.

To sum up, the calculation of the charge-carrier density not only lacks a clear definition but also is very error prone, both in simulations and in the experiment. For these reasons, we refrain from using the charge-carrier density in our calculations and use the slope of the chemical capacitance $C_\mu$ instead to estimate the Urbach energy $E_U$.

# III. Modeling of the Organic Solar Cells

For the verification of the methods used in this work and for further understanding of the experimental results, we used electro-optical simulations. The drift-diffusion simulations were performed with Matlab and a Solar Cell Capacitance Simulator, in short SCAPS, developed by Prof. Marc Burgelman and his coworkers at the University of Gent.[9, 10] It was originally developed for thin film technologies such as $Cu(In,Ga)Se_2$ or CdTe solar cells[11] but has already been successfully applied to organic solar cells as well.[6, 12]

To realize the bulk heterojunction of organic solar cells, we used the effective medium approximation which simplifies the donor-acceptor network as one semiconducting material.[9] Here, the lowest unoccupied molecular orbital (LUMO) of the acceptor is treated as the conduction band and the highest occupied molecular orbital (HOMO) of the donor as the valence band. The energetic disorder of the organic semiconductor is implemented with the use of exponential band tails. In the simulations, the exponential density of tail states is discretized to a certain number of energy levels. The conduction-band tail states are treated as acceptor-like defects which are negatively charged when occupied by an electron. Vice versa, the donor-like valence band tail states are positively charged when occupied by a hole. This charge trapped in the defect states also contributes to the total charge inside the solar cell in the drift-diffusion simulations. All simulation parameters are summarized in Table S1.

Since SCAPS is not capable of computing more complex optical models, the generation rate is imported and previously generated with the Advanced Semiconductor Analysis (ASA) software.[9] It was developed by the group of Prof. Miro Zeeman at Technical University of Delft and is commercially available under https://asa.ewi.tudelft.nl/ (accessed 19/07/2022). Since our aim is not to fit experimental data but instead perform rather generic simulations, we used optical data available from Reference [13]. The stack modelled herein consists of a 0.7 mm thick glass substrate, 130 nm of indium tin oxide, 18 nm of PEDOT:PSS as hole transport layer, an active layer with the blend PM6:Y6, 20 nm PDINO as electron-transport layer and 80 nm aluminum as cathode material. The light source used for optical simulations is the Airmass 1.5G solar spectrum.

**Table S1.** Parameters used in SCAPS to simulate admittance measurements on organic solar cells with energetic disorder if not stated otherwise. The first column of values model a solar cell with different severity of tail states. The second set of parameters describe a device with two overlapping tails with the values for the secondary tail written parentheses.

| Parameter | Figure 1+ Figure 3+ Figure 8 +Figure 9 | Figure 7 |
|---|---|---|
| Active layer thickness $d$ (nm) | 100 | 100 |
| Effective density of states conduction band/valence band $N_{CB/VB}$ (cm$^{-3}$) | $10^{19}$ | $10^{19}$ |
| Energy gap $E_{gap}$ (eV) | 1.33 | 1.33 |
| Injection barrier front contact $\phi_{bf}$ (eV) | 0.1 | 0.1 |
| Injection barrier back contact $\phi_{bb}$ (eV) | 0.1 | 0.1 |
| Electron mobility $\mu_n$ (cm$^2$V$^{-1}$s$^{-1}$) | $5\times10^{-4}$ | $2\times10^{-4}$ |
| Hole mobility $\mu_p$ (cm$^2$V$^{-1}$s$^{-1}$) | $5\times10^{-4}$ | $2\times10^{-4}$ |
| Relative dielectric permittivity $\varepsilon_r$ | 4 | 4 |
| Direct recombination coefficient $k_{dir}$ (cm$^3$ s$^{-1}$) | $10^{-12}$ | $10^{-12}$ |
| Effective density of conduction/valence band tail states $N_{CBT}/N_{VBT}$ (cm$^{-3}$eV$^{-1}$) | $10^{19}$ | $10^{19}(10^{16})$ |
| Urbach energy $E_U$ (meV) | 30/55/80 | 26 (30/55/80) |
| VBT hole/CBT electron capture coefficient $\beta_1$ (cm$^3$s$^{-1}$) | $10^{-8}$ | $10^{-8}$ |
| VBT electron/CBT hole capture coefficient $\beta_2$ (cm$^3$s$^{-1}$) | $10^{-11}$ | $10^{-11}(10^{-9})$ |
| Frequency (kHz) | 10 | 10 |

## IV. Estimation of the Chemical Capacitance

Although the solar cell in total is neutrally charged, the separation of charge carriers evokes a capacitance $C$ inside the device which is schematically shown in Figure S2a. On the one hand, there is a high number of charge carriers on the electrodes itself contributing to the electrode capacitance $C_\sigma$. On the other hand, the electrons and holes inside the solar cell contribute to the chemical capacitance $C_\mu$. The combination of both, the total capacitance $C$, can be extracted from measurements with an alternating voltage. The output data of these measurements, the complex admittance $Y$ of the solar cell, is given by

$$Y = \text{Re}[Y] + i\,\text{Im}[Y]. \tag{S9}$$

To extract the capacitance from this admittance data, one has to assume an equivalent circuit of the solar cell. To enable analytical solving of the equivalent circuit for the capacitance $C$, we assumed a simple RC-circuit as depicted in Figure S2b. In this model, the admittance can be written as

$$Y = \left(i\omega C + \frac{1}{R}\right), \tag{S10}$$

where $\omega$ is the angular frequency and $R$ a parallel resistance. Solving Equation (S9) and Equation (S10) for $C$ yields

$$C = \frac{1}{\omega}\text{Im}[Y]. \tag{S11}$$

However, this total capacitance $C$ still includes the electrode capacitance $C_\sigma$. To best model $C_\mu$, we subtract the capacitance at a working point, where the active layer is ideally fully depleted, from the capacitance at the targeted working point. This approach is also frequently used in literature.[2-5, 7] However, there are multiple options for working points for this calculation as one can measure the solar cell's admittance at different frequencies, voltages and illumination conditions. To find the most accurate approach, we simulated admittance spectroscopy measurements with SCAPS (see Section III). The resulting capacitances for the working points relevant in this work, namely at different voltages in the dark and under illumination at open circuit, are shown in Figure S2c and S2d. Generally, we are interested in the capacitance $C_{d,low}$ and $C_{ill,low}$ at low frequencies in the dark and under illumination, respectively, since it includes all free charge carriers. The simulations also allow us to calculate the actual electrode capacitance

$$C_\sigma = \varepsilon \frac{\partial E_{c,a}}{\partial V} \tag{S12}$$

with the dielectric permittivity $\varepsilon$ and the electric field $E_{c,a}$ at the cathode or anode. This simulated $C_\sigma$ is shown with the dotted, dark blue line. So, the difference between the capacitance at low frequencies $C_{low}$ and the electrode capacitance $C_\sigma$ in dark blue is the actual chemical capacitance $C_\mu$ in the simulations. Experimentally, the electrode capacitance $C_\sigma$ is not accessible. Hence, we need to explore different options of replacing the electrode capacitance $C_\sigma$ with a capacitance that can be measured to find the best approximation. Here, we can use the dark capacitance either at high or low frequencies and at reverse bias or voltage dependent. Figure S2c and S2d illustrate that in our case, the difference between the voltage-dependent capacitance at low frequencies and the dark capacitance $C_{d,high}$ at high frequencies and reverse bias best resembles the actual chemical capacitance $C_\mu$. Therefore, when speaking of the chemical capacitance $C_\mu$ in the scope of this work, we used the difference $C_\mu = C_{low} - C_{d,high}(-3\text{ V})$ with capacitances calculated with a RC-circuit. Still, we want to emphasize that this approach is simply an approximation that uses $C_{d,high}(-3\text{ V})$, also called the geometric capacitance $C_{geo}$, instead of the actual electrode capacitance $C_\sigma$.

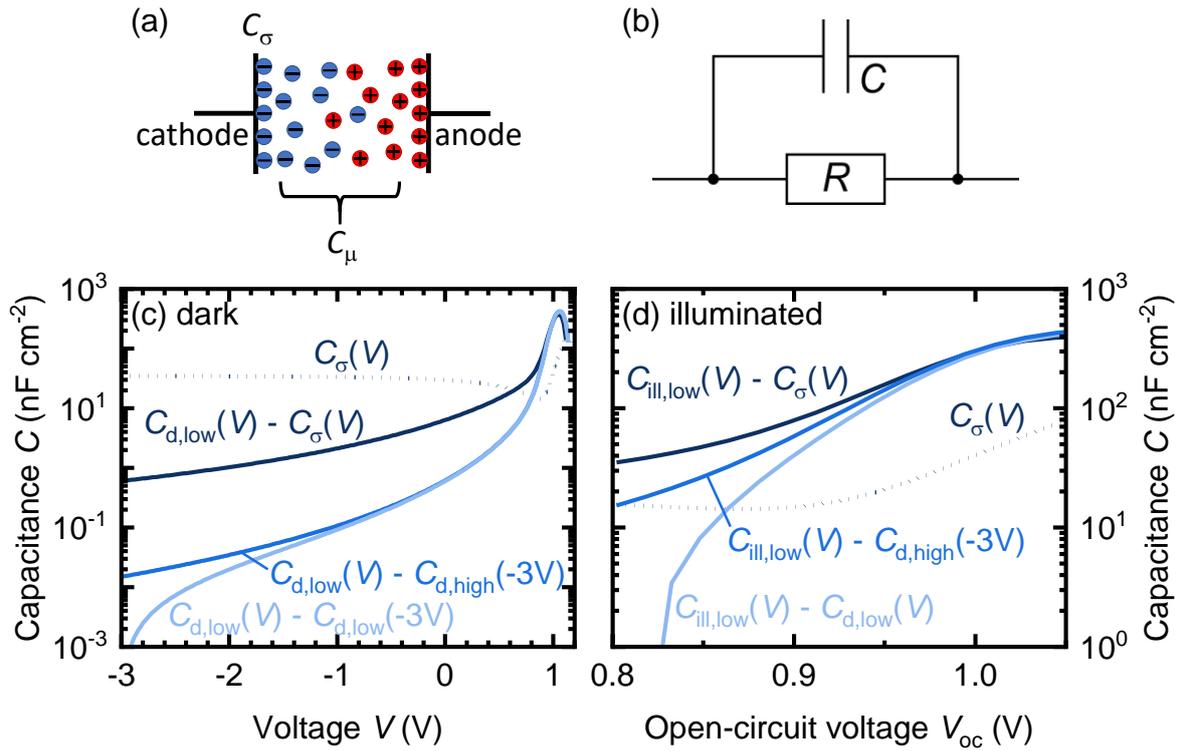

**Figure S2.** (a) Schematic of the capacitance inside a solar cell consisting of the electrode capacitance $C_\sigma$ and the chemical capacitance $C_\mu$. (b) Equivalent circuit consisting of a resistance $R$ and a capacitance $C$ that is used to calculate the capacitance from admittance data. (c) Capacitances in the dark and (d) under illumination as a function of voltage. The actual chemical capacitance $C_\mu$ is given by the difference $C_{low} - C_\sigma$ which is best replicated by $C_\mu = C_{low} - C_{d,high}(-3\ \text{V})$. The indices "low" and "high" indicate the frequency and "d" and "ill" stand for dark and illumination, respectively.

V. **Organic Solar Cell Fabrication**

For this study, we fabricated organic solar cells based on the bulk heterojunction PffBT4T-2OD:EH-IDTBR and PBDB-TF-T1:BTP-4F-12. We used glass substrates on which indium tin oxide (ITO, ca. 130 nm) was predeposited by PsiOTec Ltd. We subsequently cleaned the substrates under ultra-sonication in water, acetone and isopropyl alcohol. After drying on a hotplate at 100°C for 10 min, the substrates were treated with oxygen plasma (Diener Zepto, 13.56 MHz, 50 W) for 10 min. As a hole transport layer, we used PEDOT:PSS Al4083 by Heraeus Clevios which we treated with ultra-sound for 15 min under permanent cooling and filtered with a 0.45 µm-PVDF filter prior to spincoating. Then, 60 µl PEDOT:PSS were deposited on the ITO substrates, spincoated for 30 s at 4000 rpm and annealed on a hotplate at 150°C for 15 min, forming a ca. 30 nm film. Next, the substrates were transferred into a nitrogen filled glovebox.

For the fabrication of the PffBT4T-2OD:EH-IDTBR layer, we dissolved the donor and acceptor in a 1:1 ratio with a total concentration of 13 mg/ml in chlorobenzene and stirred over night at 100°C. Before spincoating, the substrate, the chuck of the spincoater and the tips of the pipette were preheated to 100°C. Then, 40 µl of the blend solution were spincoated at 2200 rpm for 40 s. The layer was dried at 80°C for 5 min, afterward.

Alternatively, we used PBDB-TF-T1:BTP-4F-12 as the active layer materials. Here the donor-acceptor ratio is 1:1.2 at a concentration of 18 mg/ml in o-xylene. The solution was stirred over night. One hour prior to spincoating, 0.5 % of 1,8-diiodoctane were added. We then deposited 40 µl of the solution at 2200 rpm and spun for 30 s. Next, we dried the substrates at 100°C for 10 min.

For the electron-transport layer, we used PFN-Br in a 0.5 mg/ml solution in methanol. PFN-Br, as well as the other organic semiconductors used herein, was purchased from 1-Material. Before spincoating, we filtered the PFN-Br solution with a 0.45 µm-PVDF filter. Then, we spincoated 40 µl solution at 2500 rpm for 30 s on top of the active layer, creating a 10 nm-layer.

Finally, we transferred the substrates into an evaporation chamber and deposited approximately 100 nm of silver with a shadow mask at a pressure of around $5\times10^{-7}$ mbar. The overlap of the silver cathode and the ITO anode defines the solar cell area to 6 mm².

The samples that we prepared for the photothermal deflection spectroscopy measurements were fabricated on glass substrates from the same solution and with the same parameters as the active layers of the solar cells. These substrates were previously cleaned with the same steps in the ultra-sound bath as the ITO substrates.

## VI. Solar-cell characterization

For characterization, we encapsulated the organic solar cells in measurement boxes to protect them from ambient air. The glass of the encapsulation therefore adds additional transmission and reflection losses to the photocurrent measured. A solar simulator by the brand LOT Quantum Design provides the solar spectrum for the illuminated measurements. We performed the current-voltage and admittance measurements with a potentiostat Interface 1000 by the company Gamry Instruments. The voltage range used in all experiments was -3 V to 0.9 V for the cells based on PBDB-TF-T1:BTP-4F-12 and -3 V to 1.2 V for PffBT4T-2OD:EH-IDTBR-based solar cells. We used a frequency of 10 kHz for all admittance measurements in the dark and under illumination and performed an additional measurement in the dark at 1 MHz to extract the geometric capacitance.

The Fourier-transform photocurrent spectroscopy (FTPS) measurements were carried out as reported in Reference [14]. The measurement setup consists of a FTIR spectrometer (Bruker Vertex 80v) additionally equipped with an external halogen lamp (Osram HLX100W) as light source and a low noise current amplifier (Femto DLPCA-200) connected to the solar cell. The settings for the FTIR were a mirror speed of 2.5 kHz and a resolution of 12 cm$^{-1}$. Under illumination, the photocurrent of the solar cells was amplified and then the output voltage of the current amplifier is used by software of the FTIR to create the photocurrent spectrum. To increase the dynamic range, different edge filters were placed in the beam path for additional measurements. The filters suppress the high signal above their specific wavelengths, namely above 830 nm and 1000 nm, to increase the dynamic range around the optical bandgap. For each of the filters and without the filters, at least 500 scans were carried out to decrease the signal-to-noise ratio. Finally, the measurements with the different filters were assembled to create the full FTPS spectrum.

For the measurements of photothermal deflection spectroscopy (PDS), the samples were placed on a sample holder in a cuvette filled by 2,2,3,3,4,4,5-heptafluoro-5-(1,1,2,2,3,3,4,4,4-

nonafluorobutyl)tetrahydrofuran (FC-75) that was purchased from Alfa Aesar. A halogen lamp (Osram HLX100W) was used for the optical excitation of the organic layer. A monochromator (Newport Cornerstone CS260) with filters (OG590, RG715) facilitates the sweep from high to low energies. To probe the change in refractive index of the liquid, we use a Uniphase diode laser system with a wavelength of 632 nm. A quadrant diode (Hamamatsu Photonics Quadrant Detector S1557) then measures the deflection of the laser beam. The signal is then amplified by a lock-in amplifier (Stanford research systems SR850) and collected by a custom-made software. For the calculation of the absorption coefficient from the raw data, a refractive index of 2.5 was assumed and the data was corrected by the phase shift caused by the glass substrate.

For further details and theoretical background on the optical measurement techniques FTPS and PDS, we refer to reference [15].

## VII. Determining the Effective Capacitance with an Internal Series Resistance

As described in Section IV, in this work, we use a simple RC-circuit for the calculation of the capacitance from the admittance $Y$. Hence, the capacitance $C$ is simply given by $C = \text{Im}[Y]/\omega$ with the angular frequency $\omega$. However, if the circuit is more complicated, the imaginary part of the admittance will yield an effective capacitance $C_{\text{eff}}$ that is determined by both the resistances and capacitances in the circuit. We will focus on determining the contribution of the internal series resistance $R_s$ (which models a drop in the Fermi level inside the bulk) to this effective capacitance using two different approaches. First, we offer a model based on an alternative equivalent circuit and second, we go more in depth and present an analytical description of the problem. For both approaches, we only focus on the active layer of the device and therefore neglect the electrode capacitance, since it is in parallel to the active layer and can be accounted for by simple subtraction.

In the picture of equivalent circuits, we can add a series resistance to the *RC*-element. For the effective capacitance, we first need the admittance of the modified equivalent circuit. In our case of an additional internal series resistance $R_s$, it is given by

$$Y = \left[ R_S + \left( \frac{1}{R_{\text{rec}}} + i\omega C_\mu \right)^{-1} \right]^{-1}, \tag{S13}$$

where $R_{\text{rec}}$ is the recombination resistance. Solving this equation for the imaginary part gives

$$\text{Im}(Y) = \frac{\omega C_\mu}{\left( \frac{R_S}{R_{\text{rec}}} + 1 \right)^2 + (R_s \omega C_\mu)^2}. \tag{S14}$$

Therefore, the effective capacitance $C_{\text{eff,cir}}$, that we would measure is

$$C_{\text{eff,cir}} = \frac{\text{Im}[Y]}{\omega} = \frac{C_\mu}{\left( \frac{R_S}{R_{\text{rec}}} + 1 \right)^2 + (\omega R_s C_\mu)^2}. \tag{S15}$$

In Equation (S15), the effective capacitance will reflect the true chemical capacitance $C_\mu$ if the denominator is 1. For low frequencies where $\omega \ll \frac{1}{R_s C_\mu}$, the second term is negligibly small. So, the term containing the recombination resistance $R_{rec}$ is the most relevant. For a high recombination resistance that appears at low voltages, the term goes to 1 and $C_{eff,cir} = C_\mu$. However, for high voltages, the recombination resistance decreases to values around $R_s$ or even lower. In this case, $C_{eff,cir}$ does not increase as rapidly as $C_\mu$ because of the denominator. To test this equivalent circuit and the corresponding effective capacitance $C_{eff,cir}$, we estimate $C_{eff,cir}$ from steady-state quantities. For this purpose, we first extract $R_s$ by fitting the dark $JV$-characteristics with the diode equation

$$J(V) = J_0 \exp\left(\frac{q(V_{ext} - R_S J)}{n_{id} kT}\right). \tag{S16}$$

Here, $J_0$ is the saturation-current density, $V_{ext}$ the externally applied voltage and $n_{id}$ is the ideality factor. The resulting fits are displayed in Figure S3. The recombination resistance $R_{rec}$ describes the slope of the recombination-current density with the internal voltage. Therefore, it can be calculated from

$$R_{rec} = \frac{n_{id} kT}{J_0} \exp\left(-\frac{q(V_{ext} - R_S J)}{n_{id} kT}\right). \tag{S17}$$

The analytical estimate of the chemical capacitance $C_{\mu,ana}$ follows from $C_\mu \sim dn/dV$ as

$$C_{\mu,ana} = q \int_0^d \frac{\partial n(x)}{\partial V_{int}} dx. \tag{S18}$$

Thereby, we can approximate all three circuit elements, $R_s$, $R_{rec}$ and $C_{\mu,ana}$ from steady state quantities that we can extract from drift-diffusion simulations. So, we can compare Equation (S15) to the $\text{Im}[Y]/\omega$ that we get from AC-simulations.

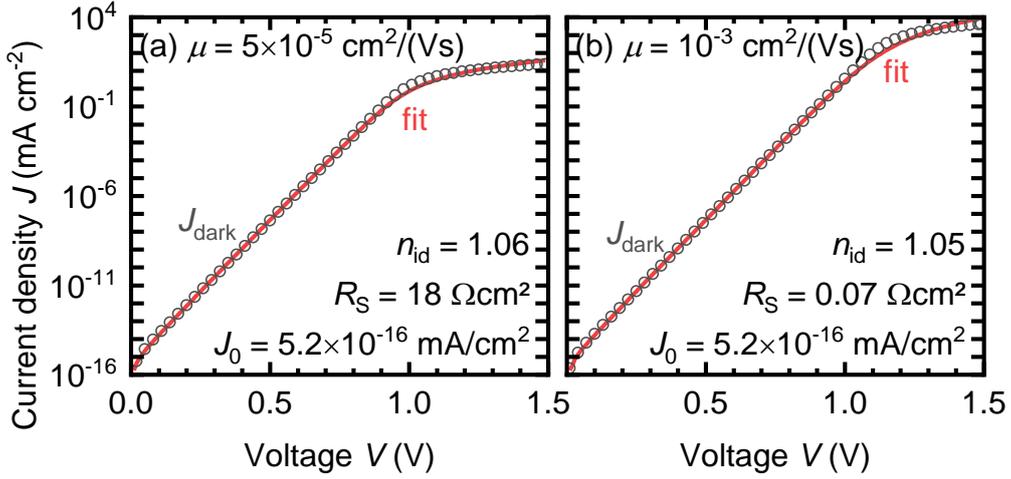

**Figure S3.** Current-density voltage curves of a simulated organic solar cell in the open circles with (a) low charge-carrier mobility $\mu$ and (b) high mobility. The red line represents a fit with the fit parameters, the ideality factor $n_\text{id}$, the series resistance $R_\text{s}$ and the saturation-current density $J_0$, displayed in the bottom right.

In the approach stated above, we simply assume the phase delay due to transport issues to be an ohmic resistance and we obtained the recombination resistance from fitted data. In the following, we present an alternative, more general, analytical approach to the calculation of the effective capacitance of such a system. This analytical derivation allows us to estimate the admittance from steady state data by using the continuity equation

$$\frac{\partial n(x,t)}{\partial t} = \frac{1}{q}\frac{\partial J_\text{n}(x,t)}{\partial x} - U(x,t) \, . \tag{S19}$$

For alternating currents, the charge-carrier density is $n(x,t) = \bar{n}(x) + \hat{n}(x,t)$ with the bar symbolizing the steady-state quantity and the hat on the carrier density $\hat{n}$ represents the alternating contribution. Similarly, the electron-current density $J_\text{n}(x,t) = \bar{J}_\text{n}(x) + \hat{J}_\text{n}(x,t)$ and the recombination rate $U(x,t) = \bar{U}(x) + \hat{U}(x,t)$. So, Equation (S19) becomes

$$\frac{\partial \bar{n}(x)+\hat{n}(x,t)}{\partial t} = \frac{1}{q}\frac{\partial \bar{J}_\text{n}(x)+\hat{J}_\text{n}(x,t)}{\partial x} - (\bar{U}(x) + \hat{U}(x,t)) \, . \tag{S20}$$

Since the continuity equation holds in steady state, the time-dependent contribution can be treated separately.

$$\frac{\partial \hat{n}(x,t)}{\partial t} = \frac{1}{q}\frac{\partial \hat{J}_\text{n}(x,t)}{\partial x} - \hat{U}(x,t) \, . \tag{S21}$$

Using Laplace transformation $\tilde{f}(\omega) = \int_0^\infty \hat{f}(t)\exp(-i\omega t)\mathrm{d}t$ for a function $f$ to go from time (hat) into frequency domain (tilde), we get

$$i\omega \tilde{n}(x,\omega) = \frac{1}{q}\frac{\partial \widetilde{J_n}(x,\omega)}{\partial x} - \widetilde{U}(x,\omega). \tag{S22}$$

Next, we rearrange the terms and integrate over the entire thickness $d$ of the active layer.

$$\int_0^d \frac{\partial \widetilde{J_n}(x,\omega)}{\partial x}\mathrm{d}x = q\int_0^d [i\omega \tilde{n}(x,\omega) + \widetilde{U}(x,\omega)]\mathrm{d}x \tag{S23}$$

$$\widetilde{J_n}(d,\omega) - \widetilde{J_n}(0,\omega) = q\int_0^d [i\omega \tilde{n}(x,\omega) + \widetilde{U}(x,\omega)]\mathrm{d}x \tag{S24}$$

Assuming a blocking contact for electrons ($\widetilde{J_n}(d,\omega) = 0$), we get the admittance as

$$Y = \frac{-\widetilde{J_n}(0,\omega)}{\widetilde{V}_{\text{ext}}(\omega)} = \frac{q}{\widetilde{V}_{\text{ext}}(\omega)}\int_0^d [i\omega \tilde{n}(x,\omega) + \widetilde{U}(x,\omega)]\mathrm{d}x, \tag{S25}$$

where $\widetilde{V}_{\text{ext}}(\omega)$ is the frequency-domain alternating voltage. In a first order approximation, the carrier density can be written as

$$\tilde{n}(x,\omega) = \left(\frac{\partial \bar{n}(x)}{\partial \bar{V}_{\text{int}}(x)}\right)\widetilde{V}_{\text{int}}(x,\omega) \tag{S26}$$

where $\bar{V}_{\text{int}}$ is the quasi Fermi level splitting in the bulk and the recombination rate is

$$\widetilde{U}(x,\omega) = \left(\frac{\partial \bar{U}(x)}{\partial \bar{V}_{\text{int}}(x)}\right)\widetilde{V}_{\text{int}}(x,\omega). \tag{S27}$$

Thereby, the admittance can be expressed as

$$Y(\omega) = q\int_0^d \frac{\widetilde{V}_{\text{int}}(x,\omega)}{\widetilde{V}_{\text{ext}}(\omega)}\left[i\omega\left(\frac{\partial \bar{n}(x)}{\partial \bar{V}_{\text{int}}(x,\omega)}\right) + \left(\frac{\partial \bar{U}(x)}{\partial \bar{V}_{\text{int}}(x,\omega)}\right)\right]\mathrm{d}x. \tag{S28}$$

The internal voltage $\widetilde{V}_{\text{int}}$ has a contribution in-phase to $\widetilde{V}_{\text{ext}}$ and an out-of-phase contribution caused by transport problems according to

$$\widetilde{V}_{\text{int}}(x,\omega) = \widetilde{V}_{\text{ext}}(\omega) - R_S(x)\tilde{J}(\omega). \tag{S29}$$

Here, the internal series resistance $R_S$ does not have to be an ohmic resistance but can be position and voltage dependent. Thus, the ratio between the internal and external voltage is given by

$$\frac{\widetilde{V}_{\text{int}}(x,\omega)}{\widetilde{V}_{\text{ext}}(\omega)} = 1 - R_S(x)Y. \tag{S30}$$

Using this expression in Equation (S28) yields

$$Y = q \int_0^d [1 - R_S(x)Y] \left[ i\omega \left( \frac{\partial \bar{n}(x)}{\partial \bar{V}_{int}(x)} \right) + \left( \frac{\partial \bar{U}(x)}{\partial \bar{V}_{int}(x)} \right) \right] dx. \tag{S31}$$

Solving Equation (S31) for the admittance gives

$$Y = \frac{q \int_0^d \left[ i\omega \left( \frac{\partial \bar{n}(x)}{\partial \bar{V}_{int}(x)} \right) + \left( \frac{\partial \bar{U}(x)}{\partial \bar{V}_{int}(x)} \right) \right] dx}{1 + q \int_0^d R_S(x) \left[ i\omega \left( \frac{\partial \bar{n}(x)}{\partial \bar{V}_{int}(x)} \right) + \left( \frac{\partial \bar{U}(x)}{\partial \bar{V}_{int}(x)} \right) \right] dx}. \tag{S32}$$

Now, the goal is to rewrite Equation (S32) in a way that enables the extraction of the imaginary part of the admittance, since it relates directly to the effective capacitance measured. For this purpose, we extend it with the complex conjugate

$$Y = \frac{\left( q \int_0^d \left[ i\omega \left( \frac{\partial \bar{n}(x)}{\partial \bar{V}_{int}(x)} \right) + \left( \frac{\partial \bar{U}(x)}{\partial \bar{V}_{int}(x)} \right) \right] dx \right) \left( 1 + q \int_0^d R_S(x) \left[ -i\omega \left( \frac{\partial \bar{n}(x)}{\partial \bar{V}_{int}(x)} \right) + \left( \frac{\partial \bar{U}(x)}{\partial \bar{V}_{int}(x)} \right) \right] dx \right)}{\left[ 1 + q \int_0^d R_S(x) \frac{\partial \bar{U}(x)}{\partial \bar{V}_{int}(x)} dx \right]^2 + \left[ q\omega \int_0^d R_S(x) \frac{\partial \bar{n}(x)}{\partial \bar{V}_{int}(x)} dx \right]^2}. \tag{S33}$$

Rearranging the numerator of Equation (S33), we get the imaginary part that correlates to the effective analytical capacitance

$$C_{\text{eff,ana}} = \frac{\text{Im}[Y]}{\omega} = \frac{q \int_0^d \frac{\partial \bar{n}(x)}{\partial \bar{V}_{int}(x)} dx}{\left[ 1 + q \int_0^d R_S(x) \frac{\partial \bar{U}(x)}{\partial \bar{V}_{int}(x)} dx \right]^2 + \left[ q\omega \int_0^d R_S(x) \frac{\partial \bar{n}(x)}{\partial \bar{V}_{int}(x)} dx \right]^2}, \tag{S34}$$

Comparing Equation (S34) with the effective capacitance from the equivalent circuit in Equation (S15), there is a striking resemblance. The numerator in Equation (S34) is the chemical capacitance and thus proves the application of Equation (S18). In the denominator, $\frac{\partial \bar{U}(x')}{\partial \bar{V}_{int}(x')}$ can be viewed as a recombination resistance and $\frac{\partial \bar{n}(x')}{\partial \bar{V}_{int}(x')}$ as a chemical capacitance. Therefore, both, an analytical and an equivalent circuit approach yield a capacitance that is reduced by transport problems in a similar manner. Figure S4 shows an exemplary comparison of Equation (S15) and (S34) for different charge-carrier mobilities. Once again, as our ability to model the internal series resistance is limited, especially in its position and voltage dependence, we use the series resistance extracted from the dark *JV*-curves as a rough estimate in both Equation (S15) and (S34). It is visible that both models feature a drop in the effective capacitance similar to the one that we observe from the simulated admittance data (solid lines). The drop in the analytical model is less steep than in the equivalent-circuit model which can be attributed to the difference in calculating the recombination resistance. Therefore, we have shown both analytically and with the consideration of different equivalent circuit models that the flattening of the capacitance-voltage curves is caused by the imaginary part of the admittance containing more than just the chemical capacitance.

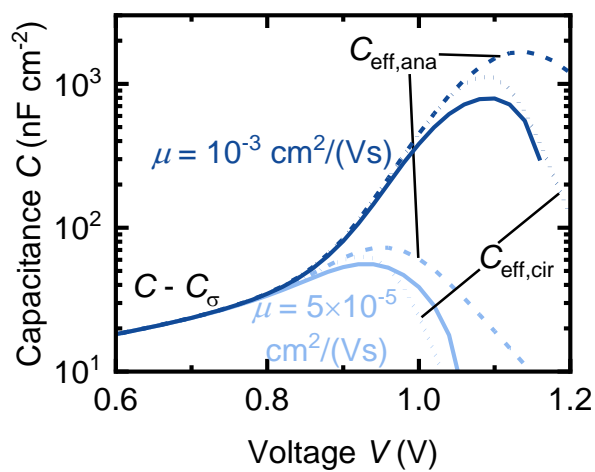

**Figure S4.** Dark capacitance-voltage simulations for organic solar cells with high and low charge-carrier mobility $\mu$. The difference between the electrode capacitance $C_\sigma$ and the capacitance $C = \text{Im}[Y]$ from simulated admittance values shows a similar drop at high voltages as our approaches to model this effective capacitance either analytically ($C_{\text{eff,ana}}$) or with an equivalent circuit ($C_{\text{eff,cir}}$).

**Table S2.** Urbach energies reported in literature for bulk-heterojunction systems. The energetic disorder was characterized by the following methods: Transient-photocurrent measurements (TPC), charge-extraction measurements (CE), Fourier-transform photocurrent spectroscopy (FTPS), photothermal deflection spectroscopy (PDS), capacitance-voltage measurements (CV), external-quantum efficiency measurements (EQE), fitting with simulations (Sim), thermally stimulated current measurements (TSC), fractional thermally stimulated current measurements (FTSC), photoabsorption spectroscopy (PAS), capacitance-frequency measurements (Cf), and space-charge limited current measurements (SCLC). The energy range is probed by optical or thermal excitation, by voltage or frequency variation or by the time-dependence of the signal. Those Urbach energies marked with * were not stated explicitly but calculated from a slope that was reported in the reference.

| Material system | Urbach Energy $E_U$ (meV) | Method | Scan type | Reference |
|---|---|---|---|---|
| BIT-4F:PC71BM | 250* | TPC | Time | [16] |
| BTR:PC71BM | 35 | CE | Voltage | [17] |
| D18:Y6 | 25.2 | FTPS | Optical | [18] |
| D18:Y6 | 24.8 | FTPS | Optical | [19] |
| D18:Y6:BTPR | 24.9 | FTPS | Optical | [18] |
| D18:Y6Se | 21.8 | FTPS | Optical | [19] |
| DF-P3HT:PC61BM | 50 | PDS | Optical | [20] |
| DPP1:PC71BM | 32* | CV | Voltage | [3] |
| DPP2:PC71BM | 36* | CV | Voltage | [3] |
| DPP3:PC71BM | 34.5* | CV | Voltage | [3] |
| DPP860:PC71BM | 29.5* | CV | Voltage | [21] |
| DPPEZnP-THD:PC61BM | 36 | EQE | Optical | [22] |
| DT-PDPP2T-TT:PC71BM | 48 | CE | Voltage | [17] |
| DTS:N2200 | 25.5* | CV | Voltage | [3] |
| DTS:PC71BM | 28.5* | CV | Voltage | [3] |
| MDMO-PPV:PC61BM | 95 | Sim | - | [23] |
| MDMO-PPV:PC61BM | 45 | EQE | Optical | [24] |
| MDMO-PPV:PC61BM | 43 | EQE | Optical | [25] |
| MEH-PPV:PC61BM | 74 | Sim | - | [23] |
| P1:PC71BM | 35.8 | PDS | Optical | [26] |
| P3HS:PC61BM | 37.2* | CE | Voltage | [27] |
| P3HT:PC61BM | 32 | EQE | Optical | [28] |
| P3HT:PC61BM | 31 | Sim | - | [23] |
| P3HT:PC61BM | 38.5* | CE | Voltage | [29] |
| P3HT:PC61BM | 53.8* | CE | Voltage | [30] |
| P3HT:PC61BM | 67.6* | CE | Voltage | [30] |
| P3HT:PC61BM | 40 | CE | Voltage | [31] |
| P3HT:PC61BM | 37 | EQE | Optical | [32] |
| P3HT:PC61BM | 35 | TPC | Time | [33] |
| P3HT:PC61BM | 65 | TPC | Time | [33] |

| Material system | Urbach Energy $E_U$ (meV) | Method | Scan type | Reference |
| --- | --- | --- | --- | --- |
| P3HT:PC61BM | 37 | EQE | Optical | [25] |
| P3HT:PC61BM | 57 | TSC | Thermal | [34] |
| P3HT:PC61BM | 42.9* | CE | Voltage | [27] |
| P3HT:PC61BM | 60 | TPC | Time | [35] |
| P3HT:PC61BM | 62* | CE | Voltage | [36] |
| P3HT:PC61BM | 30 | FTSC | Thermal | [37] |
| P3HT:PC61BM | 30 | FTSC | Thermal | [37] |
| P3HT:PC71BM | 37.1* | CV | Voltage | [21] |
| P3HT:PC71BM | 109.9 | PDS | Optical | [38] |
| P3TEA:PC71BM | 27 | PDS | Optical | [39] |
| P3TEA:SF-PDI2 | 27 | PDS | Optical | [39] |
| P3TEA:SF-PDI2 | 27 | EQE | Optical | [24] |
| PBDB-T:H1 | 24.5 | FTPS | Optical | [40] |
| PBDB-T:H2 | 24.7 | FTPS | Optical | [40] |
| PBDB-T:H3 | 23.3 | FTPS | Optical | [40] |
| PBDB-T:IDTA | 37.2 | PDS | Optical | [41] |
| PBDB-T:IDTTA | 36.5 | PDS | Optical | [41] |
| PBDB-T:IT-4F | 40.8 | FTPS | Optical | [42] |
| PBDB-T:ITIC | 78 | PAS | Optical | [43] |
| PBDB-T:ITIC | 34.4 | PDS | Optical | [44] |
| PBDB-T:ITIC | 35.2 | PDS | Optical | [44] |
| PBDB-T:ITIC | 60 | CE | Voltage | [45] |
| PBDB-T:m-4TBC-2F | 38.7 | FTPS | Optical | [46] |
| PBDB-T:N2200 | 29.6 | PDS | Optical | [47] |
| PBDB-T:o-4TBC-2F | 24.5 | FTPS | Optical | [46] |
| PBDB-T:Y6 | 41 | PAS | Optical | [43] |
| PBDB-TF:BTP-4Cl | 26.3 | EQE | Optical | [48] |
| PBDB-TF:BTP-4Cl | 23.2 | EQE | Optical | [48] |
| PBDB-TF:BTP-eC11 | 24.2 | EQE | Optical | [49] |
| PBDB-TF:BTP-eC7 | 24.3 | EQE | Optical | [49] |
| PBDB-TF:BTP-eC9 | 24.1 | EQE | Optical | [49] |
| PBDB-TF:IT-4F | 34.2 | EQE | Optical | [48] |
| PBDB-TF:IT-4F | 32.2 | EQE | Optical | [48] |
| PBDBTz-2:IT-4F | 25.5 | FTPS | Optical | [42] |
| PBDBTz-5:IT-4F | 25.3 | FTPS | Optical | [42] |
| PBDTT-DPP:PC61BM | 27 | EQE | Optical | [28] |
| PBDTT-DPP:PC61BM | 30 | Cf | Frequency | [50] |
| PBDTT-DPP:PC61BM | 30 | EQE | Optical | [50] |
| PBDTT-FTTE:m-ITIC | 75 | CE | Voltage | [45] |
| PBDTT-FTTE:O-IDTBR | 77 | CE | Voltage | [45] |
| PBDTTPD:PC61BM | 36 | EQE | Optical | [24] |
| PCDTBT:PC61BM | 47 | EQE | Optical | [28] |
| PCDTBT:PC61BM | 45 | Cf | Frequency | [50] |
| PCDTBT:PC61BM | 45 | EQE | Optical | [50] |
| PCDTBT:PC61BM | 45 | FTPS | Optical | [51] |

| Material system | Urbach Energy $E_U$ (meV) | Method | Scan type | Reference |
| --- | --- | --- | --- | --- |
| PCDTBT:PC61BM | 62* | CE | Voltage | [36] |
| PCDTBT:PC71BM | 79 | Sim | - | [23] |
| PCDTBT:PC71BM | 40.6* | CV | Voltage | [21] |
| PCDTBT:PC71BM | 75 | CE | Voltage | [17] |
| PCDTBT:PC71BM | 24 | PAS | Optical | [52] |
| PCDTBT:PC71BM | 76.7 | PAS | Optical | [52] |
| PCDTBT:PC71BM | 45 | EQE | Optical | [32] |
| PCDTBT:PC71BM | 45 | TPC | Time | [33] |
| PCDTBT:PC71BM | 45 | EQE | Optical | [53] |
| PCDTBT:PC71BM | 44.2 | EQE | Optical | [54] |
| PCDTBT:PC71BM | 44 | SCLC | Voltage | [55] |
| PCDTBT:PC71BM | 54.9 | PDS | Optical | [26] |
| PCPDTBT:PC71BM | 70.1* | CE | Voltage | [27] |
| PCPDTBT:PC71BM | 42.7 | EQE | Optical | [56] |
| PCPDTBT:PC71BM | 50.5 | EQE | Optical | [56] |
| PCPDTBT:PC71BM | 40.2 | EQE | Optical | [57] |
| PCPDTBT:PC71BM | 53.4 | EQE | Optical | [57] |
| PDCBT-2F:IT-M | 24 | EQE | Optical | [24] |
| PDPP2FT:PC71BM | 85 | TPC | Time | [35] |
| PDTTT-EFT:EH-IDTBR | 42.9* | CE | Voltage | [58] |
| PF10TBT:PC61BM | 44 | Sim | - | [23] |
| PffBT4T-2OD:EH-IDTBR | 76 | CE | Voltage | [45] |
| PffBT4T-2OD:IDFBR | 39 | PDS | Optical | [59] |
| PffBT4T-2OD:PC71BM | 31.1 | EQE | Optical | [60] |
| PffBT4T-2OD:PC71BM | 36 | PDS | Optical | [59] |
| PffBT4T-2OD:PC71BM:IDFBR | 32.8 | PDS | Optical | [59] |
| PIPCP:PC61BM | 27 | EQE | Optical | [61] |
| PIPCP:PC61BM | 27 | EQE | Optical | [62] |
| PIPCP:PC61BM | 27 | PDS | Optical | [62] |
| PIPCP:PC61BM | 27 | PDS | Optical | [63] |
| PM6:BTP-4Cl | 28 | CE | Voltage | [45] |
| PM6:BZ4F-5 | 31.2 | PDS | Optical | [64] |
| PM6:BZ4F-6 | 25.6 | PDS | Optical | [64] |
| PM6:BZ4F-7 | 22.9 | PDS | Optical | [64] |
| PM6:IDIC | 46 | EQE | Optical | [65] |
| PM6:IDIC | 26 | EQE | Optical | [65] |
| PM6:N3 | 26.2 | EQE | Optical | [66] |
| PM6:N3 | 25.4 | EQE | Optical | [67] |
| PM6:PC71BM | 46.5 | EQE | Optical | [68] |
| PM6:PC71BM | 46.7 | FTPS | Optical | [68] |
| PM6:PIDTC-T | 42.8 | EQE | Optical | [69] |
| PM6:TOBDT | 64 | EQE | Optical | [65] |
| PM6:TOBDT | 35 | EQE | Optical | [65] |
| PM6:TOBDT:IDIC | 53 | EQE | Optical | [65] |
| PM6:TOBDT:IDIC | 24 | EQE | Optical | [65] |

| Material system | Urbach Energy $E_U$ (meV) | Method | Scan type | Reference |
|---|---|---|---|---|
| PM6:Y11 | 25.6 | EQE | Optical | [68] |
| PM6:Y11 | 24.8 | FTPS | Optical | [68] |
| PM6:Y18 | 22.4 | PDS | Optical | [70] |
| PM6:Y3 | 27.7 | PDS | Optical | [70] |
| PM6:Y6 | 26.7 | EQE | Optical | [71] |
| PM6:Y6 | 30.5 | EQE | Optical | [69] |
| PM6:Y6 | 28.7 | SCLC | Voltage | [72] |
| PM6:Y6 | 27 | CE | Voltage | [45] |
| PM6:Y6:PIDTC-T | 29.4 | EQE | Optical | [69] |
| PM6:Y6:PP | 25.2 | SCLC | Voltage | [72] |
| PM7-LR:N2200 | 34.4 | EQE | Optical | [73] |
| PM7-MR:N2200 | 30.2 | EQE | Optical | [73] |
| PM7-SR:N2200 | 28.9 | EQE | Optical | [73] |
| PTB7:PC71BM | 54.6 | PDS | Optical | [74] |
| PTB7:PC71BM | 48 | EQE | Optical | [57] |
| PTB7:PC71BM | 42 | EQE | Optical | [57] |
| PTB7:PC71BM | 48.9 | PDS | Optical | [38] |
| PTB7:PC71BM:DTS | 49.9 | PDS | Optical | [74] |
| PTB7-Th:1PDI-ZnP | 49.1 | PDS | Optical | [75] |
| PTB7-Th:2PDI-ZnP | 42.7 | PDS | Optical | [75] |
| PTB7-Th:4PDI-ZnP | 38.1 | PDS | Optical | [75] |
| PTB7-Th:IEICO-4F | 39.2 | EQE | Optical | [76] |
| PTB7-Th:IOTIC | 25 | EQE | Optical | [4] |
| PTB7-Th:IOTIC-2Fa | 25 | EQE | Optical | [4] |
| PTB7-Th:IOTIC-4F | 25 | EQE | Optical | [4] |
| PTB7-Th:IOTIC-4F | 26.8 | EQE | Optical | [77] |
| PTB7-Th:ITIC | 54.4 | EQE | Optical | [78] |
| PTB7-Th:ITIC | 54 | EQE | Optical | [79] |
| PTB7-Th:PC71BM | 45 | EQE | Optical | [78] |
| PTB7-Th:PC71BM | 45 | EQE | Optical | [79] |
| PTB7-Th:PC71BM | 31.6 | EQE | Optical | [57] |
| PTB7-Th:PC71BM | 32.6 | EQE | Optical | [57] |
| PTzBI:N2200 | 32.8 | PDS | Optical | [47] |
| rr-P3HT:PC61BM | 57 | PDS | Optical | [20] |
| Si-PCPDTBT:PC71BM | 65.1* | CE | Voltage | [27] |
| TQ1:PC71BM | 35 | FTPS | Optical | [51] |

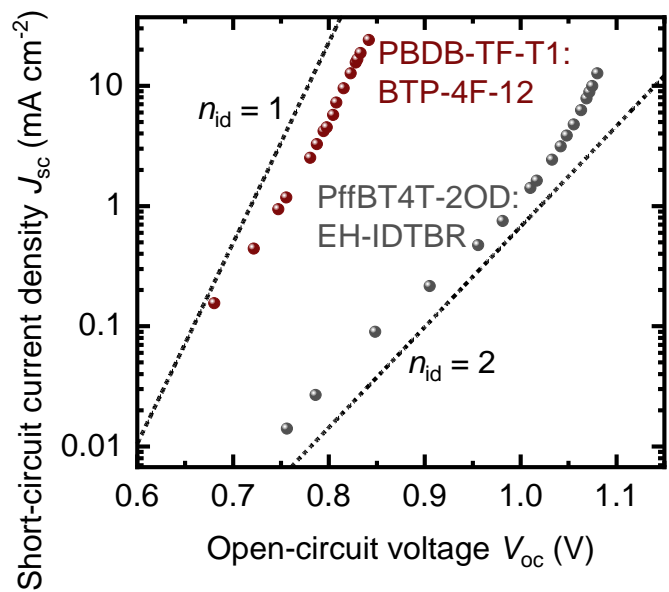

**Figure S5.** Short-circuit current density $J_{sc}$ as a function of open-circuit voltage $V_{oc}$ of the solar cells incorporating PBDB-TF-T1:BTP-4F-12 and PffBT4T-2OD:EH-IDTBR. The dashed lines indicate an ideality factor $n_{id}$ of one and two. Both devices exhibit a higher ideality factor at low light intensities indicating an increased impact of trap states further in the energy gap.

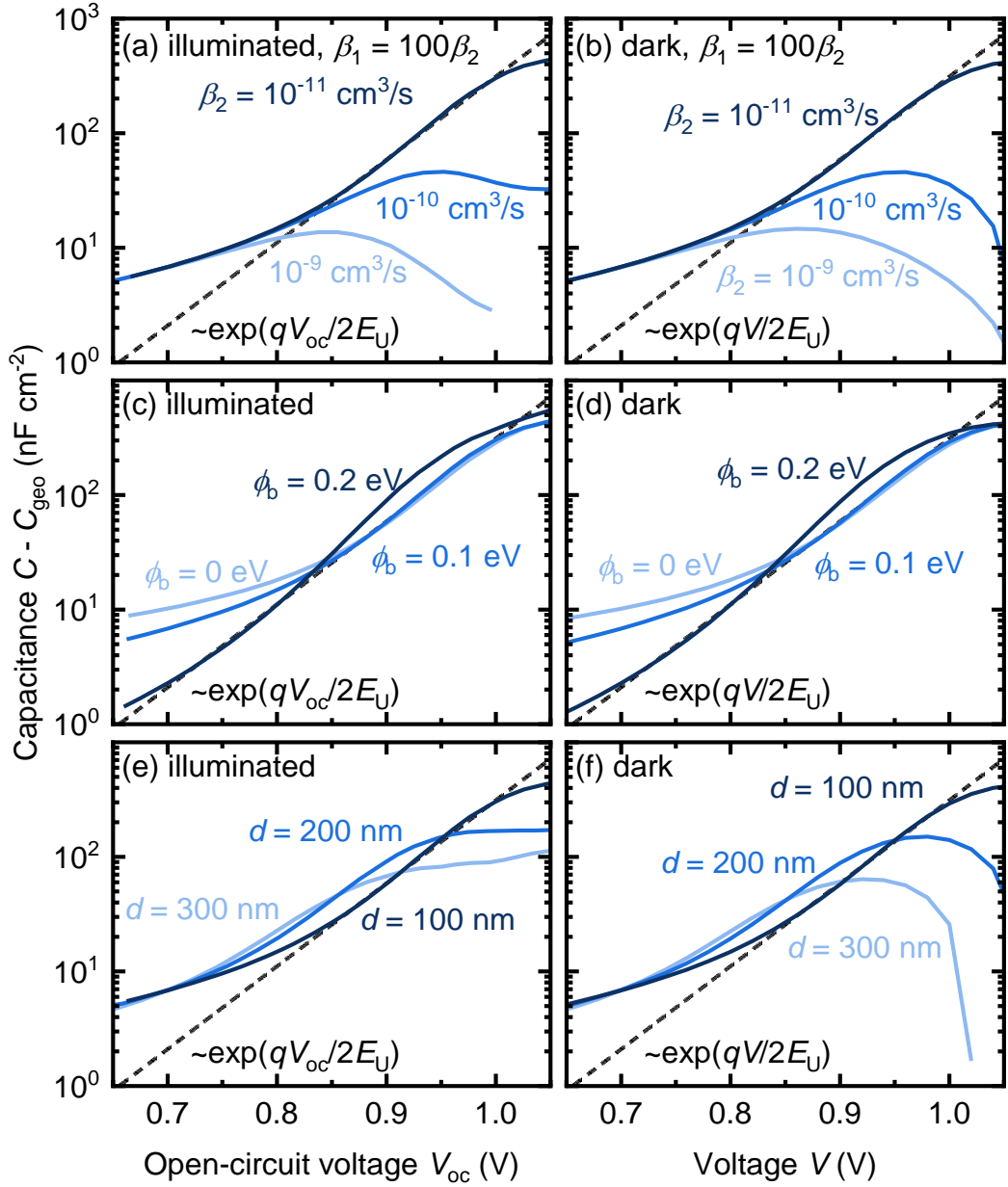

**Figure S6.** Simulated capacitance as a function of voltage (a, c, e) under illumination at open circuit and (b, d, f) in the dark. In (a) and (b), recombination via the tail states is increased by increasing the capture rates $\beta$. (c) and (d) show a variation of the injection barriers $\phi_b$ and in (e) and (f), the active layer thickness $d$ is varied from thin layers of 100 nm to thick layers of 300 nm. The dashed lines indicate the slope corresponding to the Urbach energies $E_U$ set in the simulations. In all cases, a flattening of the *CV*-curves can be observed.